\documentclass[11pt]{article}

\usepackage[utf8]{inputenc}
\usepackage[T1]{fontenc}
\usepackage{lmodern}

\usepackage[margin=1in]{geometry}
\usepackage{microtype}
\usepackage{amsmath}
\usepackage{amssymb}

\usepackage{graphicx}
\usepackage{booktabs}
\usepackage{array}
\usepackage{tabularx}
\usepackage{caption}
\graphicspath{{figures/}}

\usepackage{tikz}
\usetikzlibrary{positioning,arrows.meta,fit,calc,backgrounds}

\usepackage{xcolor}
\usepackage{listings}

\definecolor{codebg}{RGB}{247,248,250}
\definecolor{codekw}{RGB}{31,59,115}
\definecolor{codestr}{RGB}{122,68,15}
\definecolor{codecom}{RGB}{110,110,110}
\definecolor{codeframe}{RGB}{220,224,232}

\lstdefinestyle{py}{
  language=Python,
  backgroundcolor=\color{codebg},
  basicstyle=\ttfamily\footnotesize,
  keywordstyle=\color{codekw}\bfseries,
  stringstyle=\color{codestr},
  commentstyle=\color{codecom}\itshape,
  showstringspaces=false,
  breaklines=true,
  breakatwhitespace=true,
  columns=fullflexible,
  keepspaces=true,
  frame=single,
  rulecolor=\color{codeframe},
  framesep=5pt,
  xleftmargin=6pt,
  xrightmargin=2pt,
  aboveskip=0.8em,
  belowskip=0.8em,
  literate={->}{{$\rightarrow$}}1
}
\lstdefinestyle{plain}{
  backgroundcolor=\color{codebg},
  basicstyle=\ttfamily\footnotesize,
  breaklines=false,
  columns=fullflexible,
  keepspaces=true,
  frame=single,
  rulecolor=\color{codeframe},
  framesep=5pt,
  xleftmargin=6pt,
  aboveskip=0.8em,
  belowskip=0.8em
}

\usepackage[round]{natbib}
\usepackage[hidelinks]{hyperref}

\newcommand{\code}[1]{\texttt{#1}}
\newcommand{\pkg}{\textsf{pyoptexplain}}

\title{\textbf{pyoptexplain: A Python Library for\\
Post-Optimality Analysis and Explanation of Optimization Models}}
\usepackage{authblk}

\author{Hussein Fellahi\thanks{hussein.fellahi@gmail.com}}

\date{}

\begin{document}
\maketitle

% ===========================================================================
\begin{abstract}
\noindent
Optimization models are built in a variety of modeling languages and solved by a variety of solvers, but once a solution exists, the information needed to
understand it is fragmented: each solver exposes a partial, differently named set
of native diagnostics, and the modeling language has already canonicalized the
formulation the user wrote. We present \pkg{}, a practitioner-first Python
library for post-optimality analysis of optimization models that sits above this
layer. It adapts a model authored in any of five modeling front ends, namely
cvxpy, Pyomo, gurobipy, docplex, and OR-Tools, into a normalized internal
representation, solves it through a choice of backends, and answers the why and
what-if questions of an optimization decision through one uniform interface. The
design rests on two observations. First, a post-optimality quantity requested
from different backends for the same problem can come back as an exception, a structurally meaningless zero or a basis-dependent value that
disagrees across solvers, so reporting whatever one solver returns is
unreliable. \pkg{} reports a quantity only when both the representation and the chosen backend can justify it, and does not approximate unavailable information. Second, repeated scenario analysis can amortize its cost by extracting the model once and reusing a warm solver session across a batch of scenarios. \pkg{} builds a single scalable what-if interface, uniform across its modeling languages and backends and returning a certified report for every scenario, at a cost within a small constant factor of the bare solver. A reproducible computational study
substantiates both claims. Source code is available at
\url{https://github.com/h-fellahi/pyoptexplain} and installation can be done through the Python Package Index \url{https://pypi.org/project/pyoptexplain/}.
\end{abstract}

% ===========================================================================
\section{Introduction}

\pkg{} is a Python library for post-optimality analysis and explanation of
optimization models. It sits above the modeling languages used to build a model and
the solvers used to solve it, bridging several of each, so that a model authored in
one modeling framework and solved by any of several backends is analyzed through one
uniform interface. It was written for practitioners, and its purpose is to make the why
and what-if questions of an optimization decision easy to ask and to answer: which
constraints are binding, what the solution is worth at the margin, and how it would
change if a limit were loosened or a requirement removed.

In practice, this is usually where the real work lies. Mature solvers obtain a
solution reliably, but interpreting that solution is harder, because the information
one needs is scattered. Each solver exposes a partial, differently named set of native
diagnostics, and the modeling language has already canonicalized the formulation into
the solver's standard form, so those diagnostics refer to transformed variables rather
than to the objects the user wrote. The practitioner is then left to reassemble an
interpretation by hand, separately for each combination of modeling language and
solver. In an interview study of fifteen optimization developers, \citet{lawless2025}
describe the solver as a ``magical box'': powerful and trusted, but
opaque about why its answer is what it is, with validation, trust, and communication
of results recurring as practical concerns.

\pkg{} addresses this concern with a deliberately narrow scope. It adapts a model
from any supported modeling language into a normalized internal representation, and it
exposes a post-optimality analysis only when both that representation and the chosen
backend can justify it, reporting unavailable information as unavailable rather than
approximating it.

\subsection*{Contributions}

\begin{itemize}
\item \textbf{Certificate-gated honesty.} Post-optimality quantities can differ across backends, although for the same problem: requesting a dual for a constraint can sometimes yield an exception, a NaN or even a misleading 0 (see section 7.1). \pkg{} therefore only reports a quantity only when the normalized representation carries a certificate for it and the chosen backend can reliably it.
\item \textbf{Scenario analysis at scale.} Repeated solves are a key component of scenario analysis, a tool regularly used by optimization practitioners. Doing it efficiently (e.g. through solver warm starts and limited model rebuilding) becomes crucial. \pkg{} leverages the underlying solver capabilities for warm starting, and the modeling language is kept as a one-time entry point to avoid rebuilding the model at each solve. The measured per-scenario cost is similar compared to a modeling language- specific workflow for repeated solves, while returning the full certified report on every scenario. In particular, structured operations such as constraint removal are carried out through constraint deactivation to also benefit from this efficiency.
\item \textbf{Uniform coverage across modeling front ends.} We offer one analysis interface
  across five modeling front ends (cvxpy, Pyomo, gurobipy, docplex, and OR-Tools) as well as a
  direct matrix entry point. All capabilities (reports, scenario analysis etc.) are available through any of these entry points.
\end{itemize}

\subsection*{Organization}

The paper is organized as follows. Section~\ref{sec:background} reviews the surrounding literature.
Section~\ref{sec:design} describes the design philosophy and objectives, and Section~\ref{sec:architecture} the architecture
that follows from them, centered on the representation certificate and capability
gating. Section~\ref{sec:capabilities} details the analyses the library offers, and Section~\ref{sec:demos} gives examples across the supported modeling languages. Section~\ref{sec:evaluation} reports the computational study, and Section~\ref{sec:conclusion} concludes with the planned future work.

% ===========================================================================
\section{Background and related work}
\label{sec:background}

\pkg{} is a post-optimality analysis library, so it operates after a problem has
been expressed in a modeling language and solved by a solver. It therefore relates to several established bodies of work: algebraic modeling languages,
numerical solvers and their native diagnostics, classical sensitivity analysis, parametric programming, global sensitivity analysis, infeasibility analysis, and the recent literature on explainable and interpretable optimization. We review each in turn.

\textbf{Algebraic modeling languages.} Algebraic modeling languages provide a high-level surface for expressing an
optimization problem, together with a canonicalization step that converts it into the
standard form a solver accepts. AMPL \citep{fourer2003} is
the archetypal algebraic modeling language and established the separation of model from
solver that these tools share. In Python, cvxpy \citep{diamond2016} compiles
disciplined convex programs through a rewriting system \citep{agrawal2018},
and Pyomo \citep{hart2011, hart2017}
provides a general algebraic environment. Solver vendors also ship their own
modeling interfaces, namely gurobipy \citep{gurobi2026}, docplex \citep{cplex},
and OR-Tools \citep{ortools}. A wider landscape also includes lighter-weight or array-based Python modelers such as
PuLP \citep{mitchell2011}, Linopy \citep{hofmann2023} and JuMP \citep{dunning2017} in Julia.

\textbf{Solvers and their native post-optimality features.} Numerical solvers implement the algorithms that produce a solution and expose, to
varying degrees, native post-optimality information. For linear and mixed-integer
programs, HiGHS \citep{huangfu2018} is a mature open-source solver with classical
sensitivity output. For quadratic programs, OSQP \citep{stellato2020} is a widely used open-source operator-splitting solver. SCIP
\citep{achterberg2009, bolusani2024} handles mixed-integer and
mixed-integer quadratic problems. Commercial solvers such as Gurobi \citep{gurobi2026} and CPLEX \citep{cplex} provide LP, QP, and mixed-integer
solving together with sensitivity analysis and conflict or
irreducible-inconsistent-subsystem facilities. Note that the post-optimality
information these solvers expose differs in both content and interface: one returns
objective and right-hand-side ranges, another a basis or reduced costs, another an
irreducible inconsistent subsystem. Crucially, no unified cross-language tool collects
this information: each diagnostic is named and reached differently and is tied to one
stack. A practitioner reads gurobipy attributes (\code{Pi}, \code{RC},
\code{SARHSLow}/\code{SARHSUp}, \code{SAObjLow}/\code{SAObjUp}) on a Gurobi model, Pyomo
\code{Suffix} objects for duals and reduced costs, the CPLEX conflict refiner for
infeasibility, and HiGHS ranging output for sensitivity, reassembling a coherent
post-optimality picture by hand for each (modeling language, solver) pair. This
fragmentation, documented from practitioner interviews by \citet{lawless2025}, is the gap
\pkg{} is built to close, and Section~\ref{sec:scaling} measures the cost of crossing it.

\textbf{Classical sensitivity analysis.} Classical sensitivity analysis studies how the solution of a linear program responds to
changes in its data, through dual values and reduced costs and the objective and
right-hand-side ranges over which the current basis remains optimal. The standard
treatment is \citet{bertsimas1997}. \citet{wendell1985} develops the tolerance
approach, which extends ranging to simultaneous perturbations of several parameters
while the optimal basis is preserved. A recurring caveat in this literature, and one we
return to in Section~\ref{sec:evaluation}, is that the sensitivity information a solver reports can be
misleading under degeneracy and alternative optima: \citet{jansen1997}
document this and distinguish basis, optimal-partition, and optimal-value notions
of sensitivity, since a quantity read off from a single optimal basis need not
characterize the optimal value.

\textbf{Parametric and multiparametric programming.} Parametric and multiparametric programming study how an optimal solution changes as
problem data varies, partitioning the parameter space into critical regions on which the
optimal basis or active set is constant. The foundational result for linear programs is
\citet{gal1972}; the explicit-solution line, developed for constrained control, is
represented by \citet{bemporad2002}; and \citet{pistikopoulos2020}
survey the modern state of the field. These methods compute the critical regions
exactly and analytically, where \pkg{} instead samples the response empirically by
re-solving.

\textbf{Global sensitivity analysis.} Global sensitivity analysis is a distinct body of model-sensitivity work. Rather than reading
solver duals, it attributes the variance of a model output to its uncertain inputs by
sampling. Variance-based indices \citep{sobol2001}, elementary-effects screening \citep{morris1991}, and the methods collected by \citet{saltelli2008} define the field, and SALib
\citep{herman2017} is a reference Python implementation.

\textbf{Infeasibility analysis.} When a model is infeasible, infeasibility analysis identifies a minimal set of mutually
inconsistent constraints, an irreducible inconsistent subsystem. The isolation problem
was formalized by \citet{chinneck1991}, with a software approach in \citet{chinneck1994} and a book-length treatment in \citet{chinneck2008}. The production embodiments are the
irreducible inconsistent subsystem reported by Gurobi and the conflict refiner of CPLEX.

\textbf{Explainable AI.} The term explainable originates in Machine Learning, where a substantial literature
explains the predictions of learned black-box models. Feature-attribution methods such as
SHAP \citep{lundberg2017} and LIME \citep{ribeiro2016},
counterfactual explanations \citep{wachter2017}, and the Shapley
value \citep{shapley1953} from cooperative game theory are central tools in this area, and
several of them reappear as the explanation methods optimization is now borrowing.

\textbf{Explainable and interpretable optimization.} A more recent body of work applies explanation to optimization models themselves. \citet{debock2024} survey explainable AI for operational research and propose a taxonomy for
the area, and \citet{lumbreras2025} discuss why and how optimization models should
be explained. Within this area, counterfactual explanations have been developed for
discrete optimization through inverse optimization \citep{korikov2021} and for linear
optimization \citep{kurtz2026}; interpretable and data-driven
formulations build explanation into the model itself \citep{aigner2024};
and the constraint-programming community has produced step-wise
explanations of how a solution is reached \citep{bleukx2024}. A further
thread applies Shapley-value attribution to optimization, for cost allocation in linear
production games \citep{le2020} and for attributing scheduling decisions
to constraints and parameters \citep{tsuchiya2023}. \pkg{} is intended as
a foundation under this line of work: a certified post-optimality layer on which such
explanation methods can be built.

\textbf{Robust optimization.} Robust optimization speaks directly to the concern that motivates our robustness analysis
of Section~\ref{sec:perturbation}: how an optimal solution behaves when the data is uncertain or perturbed. The Robust Optimization literature
\citep{bental2009, bertsimas2011} provides the formal treatment of that concern. It builds an uncertainty set into
the model and computes a solution that remains feasible across the whole set. \pkg{}
approaches the same question from the opposite, empirical end. Rather than optimizing
against an uncertainty set in advance, it perturbs the data of an already-solved model and
re-solves, reporting how the solution, its duals, and its binding set actually move.

% ===========================================================================
\section{Design philosophy and objectives}
\label{sec:design}

\subsection{A layer above modeling and solving}

\pkg{} is not a modeling language, and it is not a solver. It occupies the layer
above both. The intuition is simple: modeling languages express a problem and
canonicalize it, solvers solve it, and \pkg{} asks what can be said about that
solve and which post-optimality analyses can be run to understand the behavior of the
problem. Our guiding principle, which we return to throughout the paper, is to let
modeling languages model and let solvers solve, and to add value strictly above that
line.

This position reflects a deliberate choice to leverage existing power rather than to
rebuild it. Modeling languages already canonicalize complex formulations, introduce
auxiliary variables, and transform constraints, and mature solvers already implement methods like
the simplex, interior-point methods, and branch-and-bound efficiently. We
therefore wrap these objects and inspect the work that they have already done. We do not attempt to be a general canonicalizer, and the internal representation
that we construct is kept intentionally limited, comprising the LP and QP matrix forms
together with a basic catch-all. We normalize a problem internally only because doing so
makes the subsequent analyses tractable and uniform.

The encouraged workflow follows directly from this. A user builds a problem in the
native modeling language of their choice, and \pkg{} adapts that object into an
internal representation suitable for analysis. Note that the native object remains the
source of truth and remains usable on its own terms: the internal representation is an
analysis surface derived from it, not a replacement for it.

This positioning has a direct payoff for scenario (repeated solve) analysis, which Section~\ref{sec:scaling}
measures. Because the modeling language is the entry point and not a per-solve dependency,
\pkg{} pays its canonicalization once, when it adapts the native object, and then drives the
solver directly across as many scenarios as the analysis requires. It thus leverages both
layers: the modeling language for translating a formulation into solver
form, and the solver for solving, scaled across repeated solves without
re-paying the translation each time. The goal is convenience and uniformity, a single layer
that unifies the modeling-language and solver landscape so that post-optimality and what-if
analyses are easy and consistent to run.

\subsection{A practitioner-first interface}
\label{sec:practitioner}

For the practitioner, the concern is understanding the solution rather than producing it,
and we organized the library around answering that concern with as little ceremony as
possible.

The interface therefore favors short, direct functions over verbose construction. The
intent is captured by a simple promise: give \pkg{} a problem in its native form,
and it returns the analyses that can be run on it. At the same time, the architecture
remains open to the power user, who can select representations explicitly, choose solver
backends, and compose experiments directly. We treat ease of use for the common case and
depth for the advanced case as complementary.

A further objective, central enough to shape the whole architecture, is honesty about
what can be supported. The library exposes only analyses that it can fully justify for
the given problem and backend, and it reports unavailable information as unavailable
rather than approximating it. Put plainly, we return a number only when we can stand
behind it. Section~\ref{sec:architecture} describes the mechanism, namely the representation certificate and
capability gating, by which this principle is enforced.

\subsection{Flexibility in the choice of solver for structured representations}

Because the structured representations are normalized and provenance-agnostic, we leave
the choice of solver for them open. A user may solve through the modeling language's own
configured solver, for example any of the solvers that cvxpy can target, or through one
of the backends that \pkg{} integrates directly, namely HiGHS, OSQP, SCIP, Gurobi,
and CPLEX. Switching backend is what unlocks different diagnostic capabilities, since a
structured representation can be sent to whichever backend exposes the information that
an analysis needs, while the user retains the freedom to solve with whatever solver they
prefer. This flexibility is a direct consequence of separating the analysis surface from
its provenance.

\subsection{A progression of post-optimality questions}
\label{sec:progression}

The analyses that the library offers can be understood as a progression, ordered by how
far they depart from the problem as originally stated.

\begin{itemize}
\item \textbf{The solve report} examines the current solution under the current inputs. It is
  static: it reports the solve status, the binding constraints, the slacks, and the
  duals, answering the question of what happened at this optimum.
\item \textbf{Robustness analysis} asks what happens when the data is perturbed slightly, for
  example by a small relative change of one or five percent in a parameter family. It
  tests whether the solution and its explanations are stable in a neighborhood of the
  original inputs.
\item \textbf{Scenario analysis} goes further and changes the model itself, by relaxing a
  constraint by a chosen amount, ablating a constraint, or altering a constraint or a
  variable, and then comparing the resulting solve with the baseline.
\end{itemize}

This progression, from observing the solve, to perturbing the data, to changing the
model, organizes the capabilities described in Section~\ref{sec:capabilities}, and it reflects the order in
which a practitioner typically interrogates a decision.

% ===========================================================================
\section{Architecture and implementation}
\label{sec:architecture}

\begin{figure}[htbp]
\centering
\begin{tikzpicture}[
  font=\small,
  node distance=8mm,
  spine/.style={rectangle, rounded corners=4pt, draw=codekw, line width=0.9pt,
                text width=8.2cm, align=center, inner sep=5pt},
  spinefill/.style={spine, fill=codebg},
  side/.style={rectangle, rounded corners=4pt, draw=codestr, line width=0.8pt,
               fill=codebg, text width=5cm, align=center, inner sep=5pt}
]
\tikzset{flow/.style={-{Latex[length=2.4mm]}, draw=codekw, line width=0.9pt}}

\node[spinefill] (model) {\textbf{user-authored model}\\[2pt]
  {\footnotesize cvxpy $\cdot$ Pyomo $\cdot$ Gurobi $\cdot$ OR-Tools $\cdot$ CPLEX $\cdot$ direct matrix}};
\node[spine, below=of model] (handle) {\textbf{ProblemHandle}\\[2pt]
  {\footnotesize preserves the native object and user-level blocks}\\[1pt]
  {\scriptsize\itshape .basic\_ / .linear\_ / .quadratic\_ / .scenario\_representation()}};
\node[spine, below=of handle] (rep) {\textbf{AnalysisRepresentation}\\[2pt]
  {\footnotesize normalized, provenance-agnostic analysis surface}};
\node[spinefill, below=of rep] (analyzer) {\textbf{Analyzer(representation{[}, backend{]})}\\[2pt]
  {\footnotesize factory $\rightarrow$ capability-gated subclass}};
\node[spine, below=of analyzer] (out) {\textbf{reports $\cdot$ experiments $\cdot$ scenarios}\\
  \textbf{grids $\cdot$ robustness $\cdot$ sensitivity}};

\node[side, right=10mm of rep] (cert) {\textbf{RepresentationCertificate}\\[2pt]
  {\footnotesize verified / asserted / unknown}\\[1pt]
  {\footnotesize block $\rightarrow$ component mappings}};

\node[side, right=10mm of analyzer, draw=gray] (gate) {\footnotesize\raggedright
  \textbf{three gating layers (\S\ref{sec:analyzer})}\\
  1. method presence\\
  2. certificate status \& capabilities\\
  3. backend support\par};

\draw[flow] (model) -- (handle);
\draw[flow] (handle) -- (rep);
\draw[flow] (rep) -- (analyzer);
\draw[flow] (analyzer) -- (out);
\draw[-{Latex[length=2.4mm]}, draw=codestr, line width=0.8pt] (rep) -- (cert);
\end{tikzpicture}
\caption{The \pkg{} pipeline. A user-authored model becomes a \code{ProblemHandle},
which exposes one or more \code{AnalysisRepresentation}s; each carries a
\code{RepresentationCertificate} recording what is verified, asserted, or unknown
together with the block-to-component mappings, and the \code{Analyzer} is constructed
over a chosen representation. Analysis is gated at the three levels of
Section~\ref{sec:analyzer}: method presence, certificate status and capabilities, and
backend support. A one-directional pipeline: each stage is chosen explicitly and never
collapsed into the next.}
\label{fig:pipeline}
\end{figure}
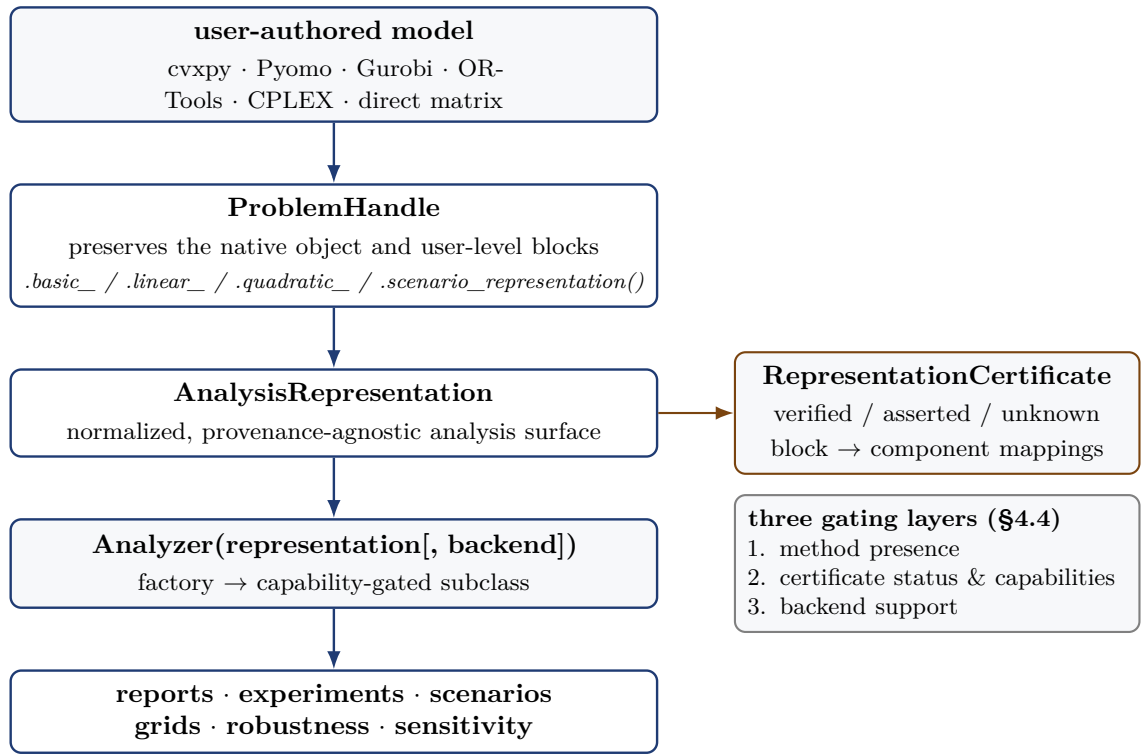

\pkg{} is organized around a single, deliberately linear pipeline that mirrors the
way a practitioner already works. A model is built in some modeling language, an analysis
surface is chosen for it, the library records what it is allowed to claim about that
surface, and only then are analyses exposed. The four stages are a \textbf{problem handle}, an
\textbf{analysis representation}, a \textbf{representation certificate}, and an \textbf{analyzer},
illustrated in Figure~\ref{fig:pipeline}.

% --- Figure 1: pipeline (TikZ recreation of fig_architecture.svg) ----------

The pipeline flows in one direction and never
collapses the stages. A handle may expose several representations with materially
different claims, the user chooses one of them explicitly, and the analyzer is
constructed over that representation rather than over the handle. This separation is what
lets the library keep its promise of Section~\ref{sec:design}: to let modeling languages model and let
solvers solve while still offering a uniform analysis surface on top. We now describe
each stage in turn.

\subsection{Problem handles}

A \textit{problem handle} is the main user-facing object. It wraps an already-built model,
preserves its high-level identity, and keeps an intentionally small protocol
(\code{pyoptexplain.core.handles.ProblemHandle}):

\begin{lstlisting}[style=py]
class ProblemHandle(Protocol):
    @property
    def constraint_blocks(self) -> list[ConstraintBlock]: ...
    def basic_representation(self) -> AnalysisRepresentation: ...
    def solve(self, *args, **kwargs) -> SolveResult: ...
\end{lstlisting}

Every handle can therefore (i) enumerate the user's constraints as \code{ConstraintBlock}s,
(ii) produce at least a basic observation surface, and (iii) solve through the native path.
Concrete handles add the representation constructors they can honestly support, such as
\code{linear\_representation()}, \code{quadratic\_representation()}, and
\code{scenario\_representation(...)}, without being required to support all of them.

Handles come in two kinds. \textbf{Direct representation handles}
(\code{LinearMatrixProblemHandle}, \code{QuadraticMatrixProblemHandle}) are entry points for a
user who writes the model explicitly in canonical matrix form; the structure is known by
construction, so no inspection is needed and the representation is trusted immediately.
\textbf{Modeling language handles} instead wrap a model already built in a supported modeling
language: each reads the language's own canonicalization, maps the recovered structure back
to the user's named constraints, and leaves the native object untouched. A third generic
\code{BasicProblemHandle} wraps an opaque solve callback for models in neither category, and
is the entry point for structures not currently supported.

The atomic unit a handle preserves is the \textbf{block}, and the constraint and variable
sides are deliberately symmetric. A \code{ConstraintBlock} (\code{core.blocks}) represents one
constraint as the user wrote it; a \code{VariableBlock} is its variable-side counterpart, one
decision variable. Both are frozen records carrying an \code{id}, a \code{name}, a \code{kind}, an
optional \code{group}, the \code{native\_object}, and \code{metadata}. A block is the user's unit of
meaning, but may map to several canonical rows once the language lowers it: \code{cp.abs(x) <= b}
is one block that canonicalizes into \code{x <= b} and \code{x >= -b}. Reports and experiments
target blocks rather than canonical rows, so a constraint named \code{labor\_capacity} appears
as \code{labor\_capacity} in every table however many rows it lowered into. Groups are
higher-level labels over blocks that also serve as removal-experiment targets, through
\code{RemoveConstraintGroup} and \code{RemoveVariableGroup}.

\subsection{Representation certificates}
\label{sec:certificates}

The \textbf{representation certificate} is the conceptual core of the design. It answers one
question: given a model from an arbitrary source, may we run a given analysis on it, and how
does a user constraint map to the canonical components the analysis actually touches? It grew
from the plug-and-play spirit of the design, where we needed to check, however the user wrote
the model, whether an analysis can safely be run.

In practice, it is created by the handle when the user calls a representation method: if the user calls \code{handle.linear\_representation()}, the handle inspects the problem it wraps (for instance checks its canonicalization) and if the inspection verifies the desired structure (here linear), the corresponding certificate is emitted.

A note on the name. A \textit{representation certificate} is not a Farkas certificate of
infeasibility or an optimality certificate. It is a trust-and-provenance record for an
analysis representation, and the codebase uses the full term consistently for that reason.

\code{RepresentationCertificate} (\code{core.certificates}) is a frozen record:

\begin{lstlisting}[style=py]
@dataclass(frozen=True)
class RepresentationCertificate:
    representation_kind: str
    problem_class: str          # "unknown" | "lp" | "milp" | "qp" | "miqp"
    status: str                 # "verified" | "asserted" | "unknown"
    source: str                 # native construction, adapter inspection, experiment, ...
    block_mappings:    Mapping[str, Sequence[RepresentationRef]]
    internal_mappings: Mapping[str, Sequence[RepresentationRef]]
    capabilities: Sequence[str]
    warnings: Sequence[str]
    limitations: Sequence[str]
    metadata: Mapping[str, Any]
\end{lstlisting}

Three parts of this record do the work.

\textbf{Status carries honesty.} A \code{verified} certificate means \pkg{} constructed or
inspected the representation through a trusted path (e.g. direct matrix specification or leveraging the canonicalization capabilities of a modeling language). An \code{asserted} certificate means the
user supplied the structural claim and mapping, which unlocks analysis without being presented
as a proof. An \code{unknown} certificate means no structural claim justifies
structure-specific analysis.

\begin{figure}[htbp]
\centering
\begin{tikzpicture}[
  font=\small,
  rowbox/.style={rectangle, rounded corners=2pt, draw=gray!60, fill=codebg,
                 minimum height=7mm, text width=4.6cm, align=left, inner sep=4pt},
  dualbox/.style={rectangle, rounded corners=2pt, draw=green!45!black, fill=green!8,
                  minimum height=7mm, minimum width=1.1cm, align=center, inner sep=3pt},
  auxbox/.style={rectangle, rounded corners=3pt, draw=codestr, fill=codebg,
                 text width=8.2cm, align=left, inner sep=4pt}
]
% user block
\node[rectangle, rounded corners=5pt, draw=codekw, line width=1pt, fill=codebg,
      text width=3.6cm, align=center, inner sep=6pt] (user)
  {\textbf{user block}\\[3pt]
   \code{l1\_budget:}\\[2pt]
   $\lVert x \rVert_1 \le \text{budget}$\\[3pt]
   {\footnotesize\itshape the unit the user\\ reasons about}};

% canonical rows
\node[rowbox, right=2.4cm of user, yshift=1.6cm] (r0) {$x_1 - t_1 \le 0$};
\node[rowbox, below=2mm of r0] (r1) {$x_2 - t_2 \le 0$};
\node[rowbox, below=2mm of r1] (r2) {$-x_1 - t_1 \le 0$};
\node[rowbox, below=2mm of r2] (r3) {$-x_2 - t_2 \le 0$};
\node[rowbox, below=2mm of r3] (r4) {$t_1 + t_2 - \text{budget} \le 0$};

% aux vars header
\node[auxbox, above=4mm of r0.north west, anchor=south west] (aux)
  {\textbf{auxiliary variables}~~$t_1, t_2$~~(\code{internal\_mappings})};

% dual column
\node[dualbox, right=4mm of r0] (d0) {1.5};
\node[dualbox, right=4mm of r1] (d1) {2.0};
\node[dualbox, right=4mm of r2] (d2) {0.5};
\node[dualbox, right=4mm of r3] (d3) {0.0};
\node[dualbox, right=4mm of r4] (d4) {2.0};
\node[above=1mm of d0, font=\footnotesize\bfseries, text=green!45!black] {verified dual};

\draw[-{Latex[length=2.4mm]}, draw=codekw, line width=1pt]
  (user.east) -- node[above, font=\scriptsize]{cvxpy} node[below, font=\scriptsize]{lowering} (r2.west);
\end{tikzpicture}
\caption{Block-to-component mapping. The single user block \code{l1\_budget}
($\lVert x \rVert_1 \le \text{budget}$) is lowered by cvxpy into auxiliary variables
$t_1, t_2$ and five generated inequality rows, each carrying its own verified component
dual. \code{block\_mappings} records this lowering, so \pkg{} reports the constraint at the
block level the user reasons about while keeping every canonical row and its dual separately
inspectable. Values are from a real cvxpy inspection run.}
\label{fig:duals}
\end{figure}
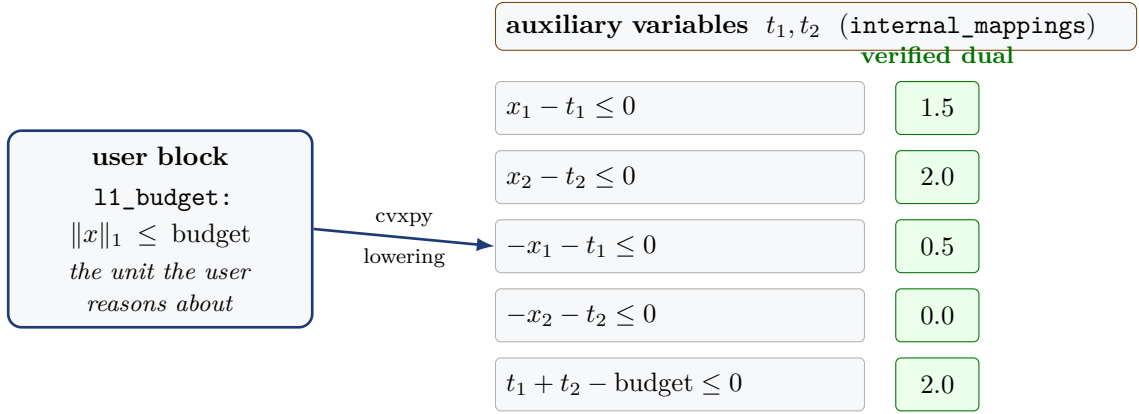

\textbf{Mappings carry the canonicalization bookkeeping.} \code{block\_mappings} associates each
user block id with a sequence of \code{RepresentationRef}s, typed pointers (\code{kind},
\code{identifier}, \code{metadata}) into the representation. The constraint \code{abs(x) <= b} may map
to two inequality rows with the roles \code{abs\_upper\_bound} and \code{abs\_lower\_bound}, while
\code{internal\_mappings} records generated components, such as auxiliary variables and epigraph
rows, that matter for provenance but are not user-facing experiment targets. A transformed
block is thus reported at its original user-level meaning while its canonical components remain
separately inspectable. Figure~\ref{fig:duals} illustrates this for an $\ell_1$-budget
constraint lowered by cvxpy.

% --- Figure 2: norm1 mapping (TikZ recreation of fig_norm1_mapping.svg) -----

\textbf{Capabilities enumerate what is on offer.} The \code{capabilities} list states, in one
place, which analyses the representation supports, so the user is told plainly what is
available and the analyzer (Section~\ref{sec:analyzer}) reads the list rather than guessing.

The gating logic is explicit. A capability must be listed in \code{capabilities}, and the
\textit{structural} subset, namely solving LP/QP/MIP/MIQP, LP and QP duals, block experiments,
classical sensitivity, scenario analysis, perturbation robustness, and integrality relaxation,
additionally requires a \code{verified} or \code{asserted} status. Non-structural capabilities
may be listed even under an \code{unknown} status. A basic representation can therefore surface
a native solver's dual as an observation, yet cannot claim LP reduced costs or classical
sensitivity.

The certificate contains no general transformation program. It records the relationship an
inspection already established rather than re-deriving it: the library is not a canonicalizer
and relies on the modeling languages for that.

\subsection{Analysis representations}
\label{sec:representations}

A handle produces an \textbf{analysis representation}, the normalized, analysis-ready object
on which the rest of the library operates. Its protocol
(\code{core.representations.AnalysisRepresentation}) is again deliberately thin:

\begin{lstlisting}[style=py]
class AnalysisRepresentation(Protocol):
    representation_kind: str
    constraint_blocks: list[ConstraintBlock]
    certificate: RepresentationCertificate
    def solve(self, options: SolveParameters | None = None) -> SolveResult: ...
\end{lstlisting}

This is the internal representation on which every analysis runs. As in
Section~\ref{sec:design}, we keep the internal modeling layer minimal and retain only what
enables analysis: the goal is to build on the modeling languages, not to duplicate them. The
representation therefore carries enough structure for the analyses we offer, and no more.

Crucially, these representations are \textbf{provenance-agnostic}. Once normalized, a
\newline \code{LinearMatrixRepresentation} carries the same analysis surface whether it came from a
direct matrix, a cvxpy inspection, or a Pyomo model. The same LP is analyzed identically
however it was built, and a verified matrix model can be re-solved on any compatible backend.

We distinguish two families: structured representations and scenario representations.

\subsubsection{Structured representations}
The structured (matrix) representations 
\begin{itemize}
    \item \code{LinearMatrixRepresentation}
    \item \code{QuadraticMatrixRepresentation}
\end{itemize} are the surface for static and local analysis: a
single solve, the classical sensitivity quantities around it, or a small perturbation of the
data. They carry the LP or QP matrix form, which is what makes the structure-specific
analyses (LP duals, reduced costs, basis status, classical sensitivity) possible at all.

\subsubsection{Scenario representations}
\textbf{Scenario representations} are the surface for analysis at scale, where one base model
is solved repeatedly under many what-if changes, as in a scenario study or a grid sweep. Here
the cost that matters is the per-case overhead paid across the whole batch, and a scenario
representation exists precisely to drive it down. There are two complementary kinds.

The \textbf{structured scenario representation}:
\begin{itemize}
    \item \code{LinearMatrixScenarioRepresentation}
    \item \code{QuadraticMatrixScenarioRepresentation}
\end{itemize}
requested through \code{scenario\_representation(extract\_structure=True)} or built from an
extracted matrix representation with \code{from\_matrix}, subclasses the structured matrix
representation, so it keeps the full matrix analysis surface but specializes it for repeated
solves:

\begin{lstlisting}[style=py]
from pyoptexplain import Analyzer, LinearMatrixScenarioRepresentation

scenario_rep = LinearMatrixScenarioRepresentation.from_matrix(handle.linear_representation())
analyzer = Analyzer(scenario_rep)          # exposes run_scenarios / explore_grid
\end{lstlisting}

Its speed comes from a \textit{lean path}. The problem is extracted into arrays once. When a
scenario case consists entirely of changes that preserve the structure and every dimension,
editing only a bound-type vector, the representation rebuilds nothing: it clones itself,
updates the touched arrays, reuses the certificate it already holds, and hands the case to a
warm solver session that stays compatible because no dimension moved. A registry of per-change
handlers decides case by case whether this applies (\code{can\_lean}) and performs the in-place
edit (\code{lean\_derive}). The lean-path changes are exactly the structure-preserving ones:

\begin{itemize}
\item a right-hand-side shift (\code{ChangeRHS}) on an inequality or equality block;
\item an inequality relaxation (\code{RelaxConstraint}), which raises the block's right-hand side;
\item an inequality removal (\code{RemoveConstraint}), realized by \textit{deactivation}: the block's rows
  are freed by setting their right-hand side to infinity, which reaches the same optimum as
  dropping the rows while leaving the matrix and every dimension intact, so that both the
  certificate and the session stay valid;
\item a variable-bound change (\code{SetVariableBounds}, \code{ChangeVariableBounds}).
\end{itemize}

Any change that alters the structure, such as relaxing or removing an equality, relaxing
integrality, or changing an objective coefficient, falls back to a full per-case derivation
that rebuilds the certificate. This split is the reason for the structured scenario
representation: the matrix form's fixed dimensions and fixed constraint matrix are exactly
what make an in-place right-hand-side edit or deactivation safe by touching a single vector,
and a warm session safe to reuse across the batch. Section~\ref{sec:changes} describes how a
typed change is dispatched onto this path, and Section~\ref{sec:scaling} measures what the lean
path saves, which is most of the per-scenario cost once a batch exceeds a handful of cases.

The \textbf{native scenario representation}:
\begin{itemize}
    \item \code{CvxpyScenarioRepresentation}
    \item \code{PyomoScenarioRepresentation}
    \item \code{GurobiScenarioRepresentation}
    \item \code{CplexScenarioRepresentation}
\end{itemize} instead keeps the original native object and mutates it in
place for each case, then restores it, touching only the blocks a scenario changes. It
complements the structured path in two ways. First, it is always available, even for problems
that cannot be extracted to a matrix at all, such as the nonlinear and conic models of
Section~\ref{sec:scaling}. Second, where the native
object exposes a richer mutation surface than a bound-vector edit, for example a live Gurobi or
CPLEX model that accepts an in-place objective-coefficient change, it can apply changes the
structured lean path would otherwise have to derive. Its price is that a front reached through
a modeling language re-enters that layer on every case. Section~\ref{sec:scenarios} records
which changes each native representation applies in place, and Section~\ref{sec:scaling}
measures the resulting trade-off against the structured path.

Together the two scenario representations are the surface used for scenario studies, grid
search over several parameters, and the optimality surfaces those grids produce.

A third representation, the \code{BasicProblemRepresentation}, is the floor. It solves and
reports whatever the native path exposes but makes no structural claim, and it solves through
the modeling language's own backend. It is the entry point for every structure that \pkg{}
does not otherwise support, keeping analyses such as constraint reports and ablation studies
available even when no structural claim can be made.

\subsection{The analyzer and capability gating}
\label{sec:analyzer}

The \code{Analyzer} is the single user-facing gateway to the analysis functions, with a
deliberately plain constructor: \code{Analyzer(representation, backend=None)}. Beneath that
surface it is a \textbf{factory}. \code{Analyzer.\_\_new\_\_} inspects the representation and
returns a representation-specific subclass, so a representation exposes only the methods it can
actually run:

\begin{lstlisting}[style=plain]
Analyzer (abstract: solve + shared reports + serialization)
|-- BasicAnalyzer              (block ablation only)
|-- MatrixAnalyzer (abstract)  (block experiments, relaxation curves,
|   |                           perturbation robustness)
|   |-- LinearMatrixAnalyzer   (lp_relaxation + classical LP sensitivity)
|   |-- QuadraticMatrixAnalyzer(qp_relaxation + quadratic objective)
|   `-- MatrixScenarioAnalyzer (structured scenarios/grids on the matrix path)
`-- ScenarioAnalyzer           (native-mutation scenarios/grids + in-place ablation)
\end{lstlisting}

Only \code{Analyzer} is exported: the subclasses are an implementation detail, though the
returned object surfaces exactly the methods its representation supports. The constructor
refuses a problem handle outright, raising a \code{TypeError} that directs the user to choose a
representation first, which enforces the pipeline's explicit-selection rule.

The base \code{Analyzer} owns the representation-independent machinery: a lazily cached base
solve (\code{solve()} and \code{result}), the block-facing reports (\code{summary}, \code{constraints},
\code{variables}, \code{duals}, \code{dual\_details}, \code{constraint\_blocks}, \code{representation\_mappings},
\code{capabilities}), and serialization (\code{to\_dict}, \code{to\_markdown}). Experiments and
scenarios solve derived representations without replacing the cached base solve, so a study
never disturbs the baseline against which it is compared.

The backend is handled per representation. Structured representations are provenance-agnostic,
so they accept any compatible backend and default sensibly when none is given; this includes
the structured scenario representation, itself a matrix representation. Basic and native
scenario representations are tied to the native solve path of the model they wrap, so an
explicit backend is meaningless and is rejected at construction. This keeps the choice of
solver consistent with what each representation can honestly support.

Capability gating then happens at three complementary levels, which together enforce the
principle that the library does not guess.

\begin{enumerate}
\item \textbf{Method presence.} An unsupported operation is absent rather than rejected at
   runtime. \code{lp\_relaxation} and \code{reduced\_costs} exist only on \code{LinearMatrixAnalyzer},
   \code{quadratic\_objective} only on the quadratic analyzer, \code{relaxation\_curve} and
   \newline \code{perturbation\_robustness} only on the matrix analyzers, and \code{run\_scenarios} and
   \code{explore\_grid} only on the structured scenario analyzer (\code{MatrixScenarioAnalyzer}) and
   the native \code{ScenarioAnalyzer}, not on the analytical matrix analyzers or a basic one.
   Calling an unsupported method raises an \code{AttributeError} rather than returning a
   misleading empty result.
\item \textbf{Certificate.} Structural analyses are gated by the representation certificate's
   status and capability list (Section~\ref{sec:certificates}): the representation must be allowed to make
   the structural claim.
\item \textbf{Backend.} The configured or class-defaulted solver backend must actually implement
   the diagnostic. HiGHS exposes LP basis and ranging information, OSQP does not expose LP
   reduced costs, SCIP claims no duals at all, and so on.
\end{enumerate}

\code{Analyzer.capabilities()} reports the effective set as the intersection of the certificate
(2) and the backend (3), while method presence (1) ensures the user never reaches a method the
intersection would deny. Every number shown is one the library can fully back, and the
unavailable contract is precise about which form an absence takes: an unavailable quantity is
returned as \code{None} when a scalar is absent, as \code{NaN} when a numeric slot in a table
has no value, or raises \code{UnsupportedFeature} when the operation itself is refused, and is
never approximated. We never invent or cast an analysis the representation and backend
cannot support.

\subsection{Experiments and scenario changes}
\label{sec:changes}

The transformations a user applies to a model (Section~\ref{sec:capabilities}) are themselves
typed objects, of two kinds that differ in how much they know about their own realization. The
distinction extends the capability gating of Section~\ref{sec:analyzer} to individual changes.

A \textbf{block experiment} (the \code{BlockExperiment} protocol in \code{analysis/experiments/})
is self-applying. It declares a \code{target\_type}, such as \code{constraint\_block} or
\code{variable\_integrality}, and implements \code{apply(representation)}, returning a derived
representation with its own certificate and transformation history. Removing or relaxing a
constraint, changing a right-hand side, removing a variable, and relaxing integrality are
block experiments. Carrying its own realization, each runs directly through
\code{Analyzer.run\_experiment(...)} and is compared against the baseline.

A \textbf{scenario change} (the \code{ScenarioChange} protocol in \code{core/changes.py}) is the
opposite: an inert, typed intent carrying only its data and a \code{describe()} method. Setting
an objective coefficient, setting variable bounds, scaling the quadratic objective, and
assigning a native parameter are scenario changes. It does not know how to apply itself: it
names what the user wants, and its \code{describe()} output exists only so a study can report
which changes produced each result.

The realization is supplied by the representation, where the two families of
Section~\ref{sec:representations} reappear. The structured (matrix) scenario representation
translates each change into a derivation that calls the matching \code{with\_...} method and
produces a derived certified representation, exactly as a block experiment does. It also
recognizes the structure- and dimension-preserving changes that edit only a bound-type vector,
a right-hand-side shift, an inequality relaxation, an inequality removal, and a variable-bound
change, for which it reuses the certificate built once at extraction and edits only the changed
entries on a warm session, falling back to a fresh derivation otherwise (relaxing or removing
an equality, relaxing integrality, changing an objective coefficient). An inequality removal
here frees the affected rows in place, setting their right-hand side to infinity, reaching the
same optimum as dropping them while keeping the certificate and session valid; the freed rows
stay in the solve report as inactive, unlike the ablation experiment that drops them entirely.
An equality cannot be loosened through its right-hand side without changing the structure, so
equality relaxation and removal stay on the fresh-derivation path. A native scenario
representation instead realizes each change by mutating the live native model in place and
restoring it. The same \code{ChangeRHS} object is therefore \textit{derived} into a new matrix
representation on one path and \textit{applied as a live mutation} on another: the change states
the intent, and each representation supplies the fitting realization. Block experiments,
already able to derive themselves, double as scenario changes on the structured path, which is
why the removal and relaxation transformations appear in both surfaces. The two protocols are
kept distinct for exactly this reason: a \code{BlockExperiment} must carry its own derivation so
it can re-certify itself for a standalone run, whereas a \code{ScenarioChange} is deliberately
inert so the representation, not the change, chooses the realization (a fresh derivation or an
in-place mutation), and the overlapping membership is the price of letting one removal or
relaxation serve both roles.

This carries capability gating down to a single change. A change is realized only on a
representation whose dispatcher can carry it out. Elsewhere it is refused with an explicit
error rather than silently approximated. A quadratic-scaling change is realized only on the
quadratic matrix path, and a change a native object cannot express in place is rejected rather
than guessed at. The principle matches the analyzer surface of Section~\ref{sec:analyzer}: \pkg{} acts only where it can fully justify the operation, and says so plainly otherwise.

\subsection{Layering and implementation discipline}

The package is layered, with imports flowing downward only. \code{core/} is a true leaf,
depending on nothing beyond the standard library, NumPy, and SciPy, and hosts the data types
and protocols (\code{blocks}, \code{certificates}, \code{changes}, \code{handles}, \code{representations},
\code{results}, \code{solvers}). \code{representations/} consumes \code{core/}; \code{handles/} consumes
\code{core/} and \code{representations/}; \code{solvers/} consumes \code{core/} and \code{representations/};
and \code{analysis/} sits on top as the orchestration layer. 

Several rules support these goals. Optional dependencies are imported lazily, so the cvxpy,
Pyomo, Gurobi, OR-Tools, and CPLEX integrations and the commercial backends are importable and
testable without their optional packages installed, keeping the base install light and the
import graph acyclic. Sparse data is preserved end to end, so SciPy CSR matrices stay sparse
through normalization, solving, and block experiments rather than being densified on entry. The
internal representations stay deliberately limited, the LP and QP matrix forms plus the basic
catch-all, so the library adds canonical modeling only where it facilitates analysis and never
drifts into becoming another modeling language or a general canonicalizer.

% ===========================================================================
\section{Analysis capabilities}
\label{sec:capabilities}

\subsection{The solve report}

The solve report is the static layer. It examines the current solution under the current
inputs and reports the solve status, the primal values, the binding constraints, the
slacks, and the duals. Every report is block-facing: it is expressed in terms of the
user's constraint and variable blocks rather than the canonical rows that a block may
have lowered into. A multi-row block is therefore reported once, at its original meaning,
and its canonical components remain available separately through the detailed dual and
mapping reports.

We report duals conservatively, and the rule is deliberately narrow. A block that maps to a
single row or a single bound receives that component's dual directly, since its marginal
value is exactly the marginal value of the block's right-hand side. A block that maps to
several rows shares one right-hand side across those rows, and the marginal value of that
right-hand side depends on how each row carries it: an inf-norm budget spreads the
right-hand side across every row, so the block marginal is the sum of the component duals,
whereas an $\ell_1$ budget lowered with auxiliary variables places it on a single carrying
row, so the block marginal is that row's dual alone. The certificate does not record this
provenance, so no scalar is sound in general. We therefore report a scalar for a multi-row
block only when every component is inactive, in which case the block dual is zero;
otherwise we withhold the scalar and direct the user to the component-level report. This is
precisely the failure the certificate exists to prevent: forwarding a single
plausible-looking aggregate would silently commit to one reading of a shared right-hand
side that the representation cannot justify.

The solve report is available for every representation, including the basic one, since it
requires only a solve and the native information that the solve exposes.

\subsection{Structural analyses}
\label{sec:structural}

The structured representations support analyses that depend on a known LP or QP structure.
For a linear representation these are the classical sensitivity quantities, namely reduced
costs, basis status, right-hand-side ranges, and objective-coefficient ranges. For a
quadratic representation they include the effective quadratic objective. A mixed-integer
representation additionally exposes its continuous relaxation, the LP relaxation of a MILP
or the QP relaxation of an MIQP, on which the relaxation duals and, in the linear case,
the classical sensitivity quantities can then be read.

Note that capability gating makes classical sensitivity analysis available only for linear representations solved by a backend
that exposes a simplex basis and ranging information, which among the integrated backends
are HiGHS, Gurobi, and CPLEX. It is never offered for a quadratic, mixed-integer, or basic
representation.

\subsection{Experiments and typed changes}
\label{sec:experiments}

An experiment is a typed transformation applied to a representation, after which the
derived solve is compared against the baseline. Each transformation is a small, named
object that records its own description, so a study reports exactly which changes produced
each result. We group the transformations below by what they act on.

Constraint-side transformations:

\begin{itemize}
\item \code{RemoveConstraint} ablates one constraint block, measuring the effect of dropping that
  requirement on the optimum.
\item \code{RemoveConstraintGroup} ablates every block in a named group at once.
\item \code{RelaxConstraint} loosens a constraint by a chosen nonnegative amount, in the direction
  implied by its sense (raising an inequality's right-hand side or widening an equality
  band), measuring the value of slackening it.
\item \code{ChangeRHS} sets or shifts a constraint block's right-hand side by a signed amount,
  which may tighten the constraint as well as loosen it.
\end{itemize}

Variable-side transformations:

\begin{itemize}
\item \code{RemoveVariable} fixes a decision variable to a supplied value and folds its linear and
  quadratic contributions into the objective constant and the right-hand sides. This can
  change the problem class, for example turning a QP into an LP when the removed variable
  carried the only curvature.
\item \code{RemoveVariableGroup} applies the same fixing across every member of a variable group.
\item \code{SetVariableBounds} sets a variable's lower and upper bounds, where an infinite bound is
  written as \code{None}.
\item \code{ChangeVariableBounds} shifts a variable's finite bounds by signed amounts.
\item \code{RelaxIntegrality} relaxes all or selected integer and binary variables to continuous,
  yielding the continuous relaxation. A full relaxation (turning a MILP into an LP or an MIQP into a QP) is considered a structural change and the model is then rebuilt.
\end{itemize}

Objective-side transformations:

\begin{itemize}
\item \textbf{SetObjectiveCoefficient} sets the linear objective coefficient of a variable.
\item \textbf{ChangeObjectiveCoefficient} shifts the linear objective coefficient of a variable by a
  signed amount.
\item \textbf{ScaleQuadraticObjective} scales the entire quadratic-objective matrix \code{Q} by a
  positive factor. A global rescaling of curvature models, for example, a change in the
  risk-aversion level of a portfolio problem, and a positive factor preserves the required
  semidefiniteness.
\item \textbf{ScaleQuadraticDiagonal} scales only the diagonal of \code{Q}, leaving the cross terms
  unchanged. This models a surge in individual variances with covariances held fixed. \pkg{} reports the exact lowest diagonal-scale factor that
  preserves semidefiniteness and enforces the exact semidefiniteness check on the derived
  representation.
\end{itemize}

Data transformation:

\begin{itemize}
\item \textbf{SetParameter} assigns a registered native-model parameter, such as a cvxpy
  \code{cp.Parameter} or a mutable Pyomo \code{Param}. This is the idiomatic way to vary data in
  those languages, and it can change a whole matrix such as a covariance, not only a
  scalar.
\end{itemize}

A \textbf{relaxation curve} is a convenience built on RelaxConstraint: an ordered sequence of
relaxations of one constraint over a list of amounts, solved together so that a compatible
backend reuses its solver session across the sequence.

Run individually, a transformation is a standalone experiment: the analyzer derives one
model from it and compares the derived solve against the baseline.

\begin{lstlisting}[style=py]
from pyoptexplain import RelaxConstraint, RemoveConstraint

analyzer.run_experiment(RemoveConstraint("labor"))             # ablate one block
analyzer.run_experiment(RelaxConstraint("labor", delta=10.0))  # loosen it by 10 units
analyzer.relaxation_curve("labor", deltas=[0, 5, 10, 15])      # one block over a range
\end{lstlisting}

These transformations reach the user through two surfaces. The removal, relaxation,
right-hand-side, and integrality transformations are available as standalone experiments,
run individually and compared against the baseline. The full vocabulary, including the
objective, bound, quadratic-scaling, and parameter transformations, composes into the
scenarios and grids of Section~\ref{sec:scenarios}. In every case the targets are user-level blocks,
variables, and groups, never canonical rows.

\subsection{Perturbation analysis}
\label{sec:perturbation}

Perturbation analysis asks whether the solution and its explanations are stable under small
perturbations of the data. It perturbs complete parameter families, for example the
objective vector or a right-hand-side family, by relative or absolute amounts, re-solves
each perturbed instance, and compares the objective, the primal and dual values, the
binding set, and the feasibility against the baseline.

\begin{lstlisting}[style=py]
robustness = analyzer.perturbation_robustness()
robustness.summary()   # per-case objective, primal/dual, binding-set, and feasibility deltas
\end{lstlisting}

\subsection{Scenarios and grids}
\label{sec:scenarios}

A scenario is a typed, ordered set of changes applied independently from the baseline,
solved and compared. A grid is the Cartesian product of several scenario axes, which
produces the optimality surfaces of Section~\ref{sec:progression}. 

Scenarios are reached through a scenario representation, obtained from a handle through
\code{scenario\_representation(...)}, and executed in one of two ways.  The \textbf{structured (matrix)
scenario representation}, requested with \code{scenario\_representation(extract\_structure=True)},
extracts the problem into arrays and builds its certificate once. It is used for a scenario that
preserves the problem's structure and every dimension, editing only a bound-type vector (a
right-hand-side shift, an inequality relaxation or removal, or a variable-bound change),
then edits only the changed entries of the data on a warm, reused solver session, while a
change that alters the structure falls back to a fresh derivation. 
The \textbf{native scenario
representation} instead applies each change in place on the live model, solves, and restores
the baseline, which avoids rebuilding the model for every case. Either way, \code{run\_scenarios} and \code{explore\_grid}, are exposed by a dedicated
analyzer, \code{MatrixScenarioAnalyzer} for the structured path and \code{ScenarioAnalyzer} for the
native path, and not by the analytical matrix analyzers.

\begin{lstlisting}[style=py]
from pyoptexplain import (
    Analyzer, ChangeRHS, GridAxis, LinearMatrixScenarioRepresentation, ScenarioCase,
)

scenario_analyzer = Analyzer(
    LinearMatrixScenarioRepresentation.from_matrix(handle.linear_representation())
)
scenario_analyzer.run_scenarios({                      # a set of named what-if cases
    "base": ScenarioCase(),
    "less_labor": ScenarioCase((ChangeRHS("labor", delta=-10.0),)),
})
scenario_analyzer.explore_grid([                       # the Cartesian product of axes
    GridAxis.rhs_change("labor", [10.0, 0.0, -10.0]),
])
\end{lstlisting}

The structured (matrix) scenario path is the most general: it accepts the full typed-change
vocabulary of Section~\ref{sec:experiments}, including the objective-coefficient, variable-bound, and (for quadratic
representations) quadratic-scaling changes, because any change that is not a plain
right-hand-side edit is realized by deriving a new representation. The native scenario
representations are narrower, since each can only apply the changes that its modeling
object exposes in place.

The set of changes that a native scenario representation can apply in place therefore
depends on the modeling language, because each one exposes a different mutation surface.
Table~\ref{tab:native-coverage} records this coverage.

\begin{table}[htbp]
\centering
\caption{Changes that each native scenario representation applies in place, without
rebuilding the model. Every modeling language additionally supports constraint ablation
through model rebuilding; the final column records the narrower set that supports
reversible in-place ablation.}
\label{tab:native-coverage}
\small
\begin{tabular}{lccccc}
\toprule
Modeling language & Obj.\ coefficient & Variable bounds & RHS / relaxation & Parameter & In-place ablation \\
\midrule
cvxpy  & No  & No  & No  & Yes & No \\
Pyomo  & No  & Yes & Yes & Yes & Yes \\
Gurobi & Yes & Yes & Yes & No  & Yes (ineq.) \\
CPLEX  & Yes & Yes & Yes & No  & Yes (ineq.) \\
OR-Tools$^{\dagger}$ & --- & --- & --- & --- & --- \\
\bottomrule
\end{tabular}
\\[3pt]
{\footnotesize $^{\dagger}$ Matrix path only: \code{pywraplp} has no native scenario
representation, so OR-Tools scenarios run through the extracted matrix representation
(Section~\ref{sec:scenarios}).}
\end{table}

Two patterns stand out. First, cvxpy expresses all scenario variation through registered
\code{cp.Parameter} objects, so its native scenarios accept parameter assignments alone (any
coefficient or bound study is modeled as a parameter). Second, reversible in-place constraint
ablation is available for Pyomo, unconditionally and for every block kind through constraint
deactivation, and in qualified form for Gurobi and CPLEX, which free inequalities with a
settable right-hand side by relaxing the bound to infinity: linear for both, also quadratic for
Gurobi while equalities, SOS constraints, and general blocks fall back to the rebuild of
Section~\ref{sec:experiments}. The two routes report differently: the in-place route keeps the
freed row as an inactive constraint, with slack and a zero dual, while the rebuild route drops
it entirely. cvxpy is rebuild-only by necessity: it forbids infinite parameter values, and using a large finite value instead is unstable (can be silently wrong once the
problem's scale approaches it and masking unbounded removals as finite optima).
Note on OR-Tools: \code{pywraplp} is linear-only, so OR-Tools scenarios run through the
extracted matrix representation, which serves every change type efficiently.

\subsection{Coverage by representation}

The analyses above are not implemented for every representation. Table~\ref{tab:coverage} summarizes where
each one is available across the three representation families: the basic representation,
the structured LP and QP matrix representations, and the native scenario representations.
The structured representations carry the most analysis surface, since most
structure-specific analyses require the normalized matrix form. The basic representation
supports observation and constraint removal alone. The native scenario representations
specialize in repeated solves, so they carry scenarios, grids, and in-place ablation but
not the structure-specific analyses.

\begin{table}[htbp]
\centering
\caption{Availability of each analysis by representation family.}
\label{tab:coverage}
\small
\begin{tabular}{lccc}
\toprule
Capability & Basic & Structured & Scenario \\
\midrule
Solve report (status, binding, slack, duals) & Yes & Yes & Yes \\
Constraint ablation & Yes, rebuild & Yes, derive & Yes, in-place \\
Constraint-group removal & Yes & Yes & No \\
Variable removal & No & Yes & No \\
Variable-bound change & No & Yes & Yes \\
RHS change and constraint relaxation & No & Yes & Yes \\
Objective-coefficient change & No & Yes & Yes \\
Quadratic-objective scaling (whole or diagonal) & No & Yes & No \\
Integrality relaxation & No & Yes & No \\
Relaxation curve & No & Yes & No \\
Typed scenarios & No & Yes & Yes \\
Grid exploration and optimality surfaces & No & Yes & Yes \\
Perturbation robustness & No & Yes & No \\
LP or QP relaxation of a MILP or MIQP & No & Yes & No \\
Classical LP sensitivity & No & Yes & No \\
\bottomrule
\end{tabular}
\end{table}

\textit{Notes.} In-place ablation in the scenario column follows the in-place ablation column of
Table~\ref{tab:native-coverage}: unconditional for Pyomo through deactivation, and for Gurobi and CPLEX restricted
to inequalities with a settable right-hand side, with equalities, SOS constraints, and
general blocks falling back to rebuild. Variable-bound, RHS, constraint-relaxation,
objective-coefficient, typed-scenario, and grid availability in the scenario column holds
to the extent that the modeling language supports the corresponding change in Table~\ref{tab:native-coverage}.
Quadratic-objective scaling applies to quadratic structured representations and is realized
on the matrix path; the native scenario representations do not scale the quadratic
objective in place. Classical LP sensitivity applies to linear structured representations
and is further gated by the backend, as described in Section~\ref{sec:structural}. The LP or QP relaxation
row refers to the continuous relaxation of a mixed-integer structured representation.

% ===========================================================================
\section{A tour by example}
\label{sec:demos}

\subsection{A linear program: product mix}
\label{sec:demo-lp}

Consider a factory that makes chairs and tables, each consuming labor and material, and
maximizes profit subject to the two resource capacities:
\[
\begin{aligned}
\max_{x \ge 0}\quad & 40\,x_{\text{chairs}} + 30\,x_{\text{tables}} \\
\text{subject to}\quad & 2\,x_{\text{chairs}} + x_{\text{tables}} \le 100 \quad (\text{labor}) \\
& x_{\text{chairs}} + 2\,x_{\text{tables}} \le 80 \quad (\text{material}) .
\end{aligned}
\]

The minimum input is the matrix form of the problem. A handle wraps it, a representation
is chosen, and the analyzer answers the practitioner's questions in sequence:

\begin{lstlisting}[style=py]
from pyoptexplain import (
    Analyzer, ChangeRHS, LinearMatrixProblemHandle,
    LinearMatrixScenarioRepresentation, ScenarioCase, SetObjectiveCoefficient,
)

handle = LinearMatrixProblemHandle(
    sense="max",
    c=[40.0, 30.0],
    A_ub=[[2.0, 1.0], [1.0, 2.0]],
    b_ub=[100.0, 80.0],
    bounds=[(0.0, None), (0.0, None)],
    variable_names=["chairs", "tables"],
    inequality_names=["labor", "material"],
    constraint_groups={"resources": ["labor", "material"]},
)
analyzer = Analyzer(handle.linear_representation())
\end{lstlisting}

\textbf{The optimal decision.} The optimum makes 40 chairs and 20 tables for a profit of
2200, with both the labor and material constraints binding. Because both bind, both carry a
shadow price, reported against the constraint the user named. \code{analyzer.constraints()} gives:

\begin{center}
\small
\begin{tabular}{lccc}
\toprule
name & binding & slack & dual \\
\midrule
labor    & True & 0.0 & 16.6667 \\
material & True & 0.0 & 6.6667 \\
\bottomrule
\end{tabular}
\end{center}

In words, labor is \textbf{locally} worth about 16.67 per unit and material about 6.67.

\textbf{Classical sensitivity.} A shadow price holds only while the current basis stays optimal.
\code{analyzer.rhs\_ranges()} reports how far each capacity can move before that valuation
changes, and \code{analyzer.objective\_ranges()} does the same for the profit coefficients:

\begin{center}
\small
\begin{tabular}{lccc}
\toprule
name & current\_rhs & allowable\_decrease & allowable\_increase \\
\midrule
labor    & 100.0 & 60.0 & 60.0  \\
material & 80.0  & 30.0 & 120.0 \\
\bottomrule
\end{tabular}
\end{center}

The profit coefficients carry the matching ranges, the interval over which each keeps the
current vertex optimal:

\begin{center}
\small
\begin{tabular}{lccc}
\toprule
name & current\_coefficient & allowable\_decrease & allowable\_increase \\
\midrule
chairs & 40.0 & 25.0 & 20.0 \\
tables & 30.0 & 10.0 & 50.0 \\
\bottomrule
\end{tabular}
\end{center}

The chair price may range from 15 to 60 and the table price from 20 to 80 before the optimal
product mix changes.

\textbf{The optimality curve.} The optimality curve extends the local shadow price into a global
view. \code{analyzer.relaxation\_curve("material", deltas=[0, 10, 20, 30, 40])} re-solves as the
material capacity increases:

\begin{figure}[htbp]
\centering
\includegraphics[width=0.7\textwidth]{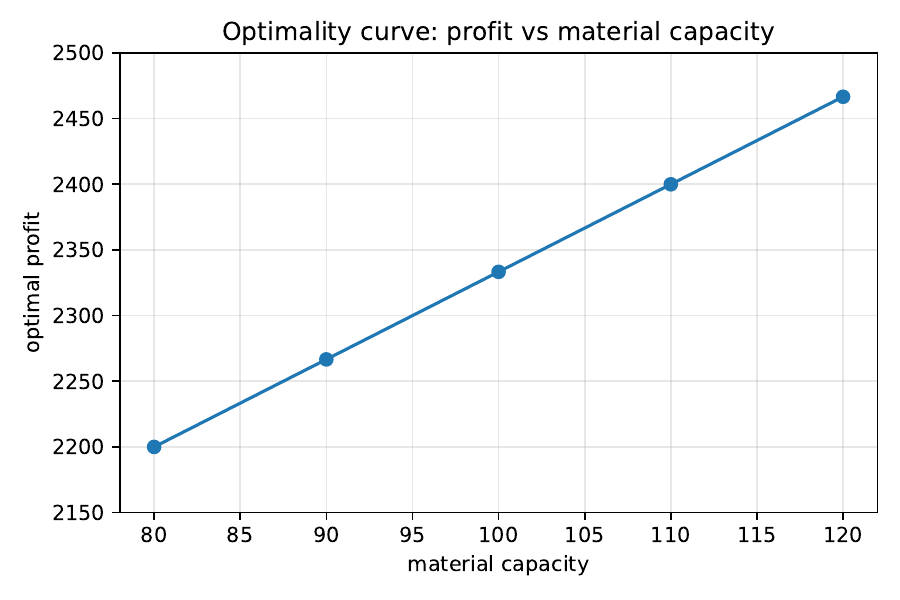}
\caption{Optimality curve for the product-mix LP: optimal profit against material capacity.
While material stays binding the profit rises linearly at its shadow price of 6.67 per unit.
Recomputed from \code{notebooks/product\_mix\_lp.ipynb}.}
\label{fig:optimality}
\end{figure}

While material stays binding, the optimal profit rises at the material shadow price of 6.67
per unit, and the curve is the planner's map of the value of additional material.

\textbf{Scenarios.} Two scenarios close the demonstration. Because they re-solve the model many
times, they run on the structured (matrix) scenario representation, obtained from the matrix
representation through \code{from\_matrix}, which builds the certificate once and reuses a warm
session across cases:

\begin{lstlisting}[style=py]
scenario_analyzer = Analyzer(
    LinearMatrixScenarioRepresentation.from_matrix(handle.linear_representation())
)
scenario_analyzer.run_scenarios({
    "base": ScenarioCase(),
    "pricier_chairs": ScenarioCase((SetObjectiveCoefficient("chairs", 55.0),)),
    "less_material": ScenarioCase((ChangeRHS("material", delta=-20.0),)),
})
\end{lstlisting}

\begin{center}
\small
\begin{tabular}{lcc}
\toprule
scenario & objective\_value & objective\_change \\
\midrule
base           & 2200.0000 & 0.0000    \\
pricier\_chairs & 2800.0000 & 600.0000  \\
less\_material  & 2066.6667 & -133.3333 \\
\bottomrule
\end{tabular}
\end{center}

Raising the profit of chairs lifts the optimum to 2800, and removing twenty units of
material lowers it to about 2066.7, a drop of exactly twenty times the material shadow
price, which confirms the local valuation against an explicit re-solve.

\subsection{A mixed-integer program: generalized assignment}
\label{sec:demo-milp}

Next, five tasks must each be assigned to one of three production lines, each line has a
limited capacity, and the total cost is minimized. With $x_{ij} \in \{0, 1\}$ indicating
task $j$ on line $i$:
\[
\begin{aligned}
\min_{x}\quad & \sum_{i}\sum_{j} c_{ij}\, x_{ij} \\
\text{subject to}\quad & \sum_{i} x_{ij} = 1 \quad \forall j \quad (\text{each task assigned once}) \\
& \sum_{j} r_{ij}\, x_{ij} \le b_i \quad \forall i \quad (\text{line capacity}) \\
& x_{ij} \in \{0, 1\} .
\end{aligned}
\]

The capacities make the problem genuinely integer, which is the point of the example:
without them the assignment constraints are totally unimodular and the relaxation would
already be integral. We build the model in matrix form, with the assignment variables
flattened into one binary vector and the two constraint families assembled into \code{A\_eq} and
\code{A\_ub}:

\begin{lstlisting}[style=py]
import numpy as np
from pyoptexplain import Analyzer, LinearMatrixProblemHandle, RelaxIntegrality

lines = ["line_A", "line_B", "line_C"]
tasks = ["T1", "T2", "T3", "T4", "T5"]
cost = np.array([[9, 12, 8, 15, 10], [11, 7, 14, 9, 13], [13, 10, 12, 8, 9]], dtype=float)
resource = np.array([[4, 5, 3, 6, 4], [5, 3, 6, 4, 5], [6, 4, 5, 3, 4]], dtype=float)
capacity = np.array([6.0, 8.0, 7.0])
n_lines, n_tasks = cost.shape

def variable_index(i, j):
    return i * n_tasks + j

# each task assigned once:  sum_i x[i, j] = 1
A_eq = np.zeros((n_tasks, n_lines * n_tasks))
for j in range(n_tasks):
    for i in range(n_lines):
        A_eq[j, variable_index(i, j)] = 1.0

# each line within capacity:  sum_j resource[i, j] x[i, j] <= capacity[i]
A_ub = np.zeros((n_lines, n_lines * n_tasks))
for i in range(n_lines):
    for j in range(n_tasks):
        A_ub[i, variable_index(i, j)] = resource[i, j]

handle = LinearMatrixProblemHandle(
    sense="min",
    c=cost.flatten(),
    variable_types=["binary"] * (n_lines * n_tasks),
    A_ub=A_ub, b_ub=capacity,
    A_eq=A_eq, b_eq=np.ones(n_tasks),
    bounds=[(0.0, 1.0)] * (n_lines * n_tasks),
    variable_names=[f"{lines[i]}->{tasks[j]}" for i in range(n_lines) for j in range(n_tasks)],
    inequality_names=[f"cap[{line}]" for line in lines],
    equality_names=[f"assign[{task}]" for task in tasks],
    constraint_groups={
        "capacities": [f"cap[{line}]" for line in lines],
        "assignments": [f"assign[{task}]" for task in tasks],
    },
)
analyzer = Analyzer(handle.linear_representation())
\end{lstlisting}

Solved by the default HiGHS backend, the minimum-cost assignment costs 43, and two of the
three line capacities bind at the optimum.

\textbf{The integrality gap.} The gap is exposed directly. \code{RelaxIntegrality} relaxes the binary
variables to the continuous interval:

\begin{lstlisting}[style=py]
analyzer.run_experiment(RelaxIntegrality()).to_frame()
\end{lstlisting}

\begin{center}
\small
\begin{tabular}{cccc}
\toprule
base\_objective\_value & objective\_value & integrality\_gap\_absolute & integrality\_gap\_relative \\
\midrule
43.0 & 41.5 & 1.5 & 0.0349 \\
\bottomrule
\end{tabular}
\end{center}

The continuous relaxation costs 41.5, a gap of 1.5 below the integer optimum, so the
relaxation is a bound rather than an achievable plan.

\textbf{Capacity valuations.} A mixed-integer optimum has no duals, so the only available
capacity valuations are read on the relaxation through \code{lp\_relaxation()}:

\begin{lstlisting}[style=py]
analyzer.lp_relaxation().duals()
\end{lstlisting}
The binding capacity in the relaxation is \code{line\_A} with a dual of $-0.5$, while the lines
binding in the integer optimum are \code{line\_B} and \code{line\_C}. Note that the binding line in the
relaxation need not be a binding line in the integer optimum, a distinction the library
keeps explicit rather than eliding.

\begin{table}[h]
\centering
\begin{tabular}{ll}
\hline
\textbf{name} & \textbf{dual} \\
\hline
cap[line\_A] & -0.5 \\
cap[line\_B] & 0.0 \\
cap[line\_C] & 0.0 \\
\hline
\end{tabular}
\end{table}

\textbf{Scenarios and ablation.} The value of a binding capacity is read by ablation. Removing the
limit on the first line and re-solving lowers the cost to 41
(\code{analyzer.ablate\_constraint("cap[line\_A]")}), so the gap between 43 and 41 is exactly what
that capacity costs at the integer optimum, the integer counterpart of a shadow price
obtained by removal rather than by a dual. Scenarios complete the picture:

\begin{lstlisting}[style=py]
scenario_analyzer = Analyzer(
    LinearMatrixScenarioRepresentation.from_matrix(handle.linear_representation())
)
scenario_analyzer.run_scenarios({
    "base": ScenarioCase(),
    "expand_line_A": ScenarioCase((ChangeRHS("cap[line_A]", delta=1.0),)),
    "pricier_T5_on_C": ScenarioCase((SetObjectiveCoefficient("line_C->T5", 12.0),)),
    "tighten_line_C": ScenarioCase((ChangeRHS("cap[line_C]", delta=-1.0),)),
})
\end{lstlisting}

\begin{center}
\small
\begin{tabular}{lccc}
\toprule
scenario & status & objective\_value & objective\_change \\
\midrule
base            & optimal    & 43.0 & 0.0  \\
expand\_line\_A   & optimal    & 41.0 & -2.0 \\
pricier\_T5\_on\_C & optimal   & 46.0 & 3.0  \\
tighten\_line\_C  & infeasible & NaN  & NaN  \\
\bottomrule
\end{tabular}
\end{center}

Expanding \code{line\_A} by one unit recovers the cheaper plan, a price increase passes straight
through, and tightening another line makes the problem infeasible, which the scenario table
surfaces as a row rather than as an error.

\textbf{An optimality surface.} A grid over two line capacities draws the surface, through
\code{explore\_grid} on the same scenario analyzer:

\begin{lstlisting}[style=py]
scenario_analyzer.explore_grid([
    GridAxis.rhs_change("cap[line_A]", [2.0, 1.0, 0.0, -1.0]),
    GridAxis.rhs_change("cap[line_C]", [1.0, 0.0, -1.0, -2.0]),
])
\end{lstlisting}

The optimal cost is piecewise constant in the capacities, as an integer optimum is in its
right-hand sides: it holds at one level while a line is loose, steps up as capacity
tightens, and becomes infeasible in the tightest corner.

\begin{figure}[htbp]
\centering
\includegraphics[width=0.7\textwidth]{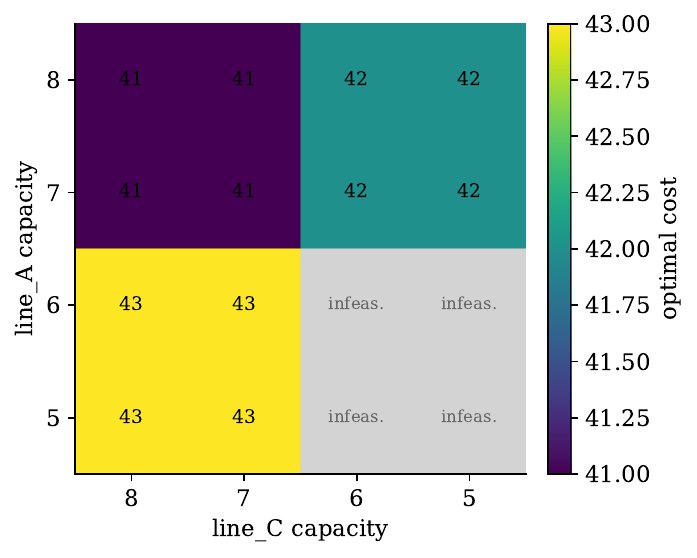}
\caption{Optimality surface for the generalized-assignment MILP: optimal cost over a grid of
two line capacities. The cost is piecewise constant, stepping up as capacity tightens, and
the tightest corner is infeasible. Recomputed from
\code{notebooks/generalized\_assignment\_milp.ipynb}.}
\label{fig:grid}
\end{figure}

\subsection{A quadratic program: mean-variance portfolio}
\label{sec:demo-qp}

For the quadratic case, capital is allocated across five assets to minimize portfolio
variance subject to being fully invested and meeting a target expected return:
\[
\begin{aligned}
\min_{w \ge 0}\quad & w^\top \Sigma w \\
\text{subject to}\quad & \mathbf{1}^\top w = 1 \quad (\text{fully invested}) \\
& \mu^\top w \ge r \quad (\text{return target}) .
\end{aligned}
\]

Risk is quadratic in the weights, so the problem is a quadratic program, built with
\code{QuadraticMatrixProblemHandle}. We pass $Q = 2\Sigma$ so the reported objective
$\tfrac{1}{2} w^\top Q w$ equals the variance, and encode the return target as an
inequality row $-\mu^\top w \le -r$:

\begin{lstlisting}[style=py]
import numpy as np
from pyoptexplain import Analyzer, QuadraticMatrixProblemHandle

assets = ["tech", "energy", "bonds", "real_estate", "gold"]
mu = np.array([0.12, 0.10, 0.04, 0.08, 0.06])    # expected annual returns
vol = np.array([0.20, 0.18, 0.06, 0.12, 0.15])   # annual volatilities
correlation = np.array([
    [1.0,  0.5, -0.2,  0.4, -0.1],
    [0.5,  1.0, -0.1,  0.4,  0.0],
    [-0.2, -0.1, 1.0,  0.0,  0.2],
    [0.4,  0.4,  0.0,  1.0, -0.1],
    [-0.1, 0.0,  0.2, -0.1,  1.0],
])
covariance = np.outer(vol, vol) * correlation
n = len(assets)
return_target = 0.08

handle = QuadraticMatrixProblemHandle(
    sense="min",
    Q=2.0 * covariance,             # objective 0.5 w' Q w = w' Sigma w = variance
    c=np.zeros(n),
    A_ub=(-mu).reshape(1, -1),      # -mu . w <= -r  encodes  mu . w >= r
    b_ub=np.array([-return_target]),
    A_eq=np.ones((1, n)),           # 1 . w = 1
    b_eq=np.array([1.0]),
    bounds=[(0.0, 1.0)] * n,        # long only
    variable_names=assets,
    inequality_names=["return_target"],
    equality_names=["fully_invested"],
)
analyzer = Analyzer(handle.quadratic_representation())
\end{lstlisting}

Solved by the default OSQP backend at an 8\% return target, the minimum-variance
portfolio is diversified across all five assets, with a realized volatility of about 0.082,
and both the budget and the return target bind.

\textbf{Constraint duals.} \code{analyzer.duals()} on the structural constraints:

\begin{center}
\small
\begin{tabular}{lc}
\toprule
name & dual \\
\midrule
return\_target  & -0.3155 \\
fully\_invested & -0.0119 \\
\bottomrule
\end{tabular}
\end{center}

The return-target dual, about $-0.32$, is the marginal variance the portfolio must accept
for one more unit of required return, which is the local slope of the efficient frontier.

\textbf{The efficient frontier.} Sweeping the return target traces the whole frontier. The return
constraint is $-\mu^\top w \le -r$, so a target $r$ is an RHS shift on the \code{return\_target}
row, swept by a grid on the scenario analyzer across
\code{targets = [0.06, 0.07, 0.08, 0.09, 0.10, 0.11]}:

\begin{lstlisting}[style=py]
scenario_analyzer = Analyzer(
    QuadraticMatrixScenarioRepresentation.from_matrix(handle.quadratic_representation())
)
scenario_analyzer.explore_grid([
    GridAxis.rhs_change("return_target", list(return_target - targets)),
])
\end{lstlisting}

\begin{figure}[htbp]
\centering
\includegraphics[width=0.7\textwidth]{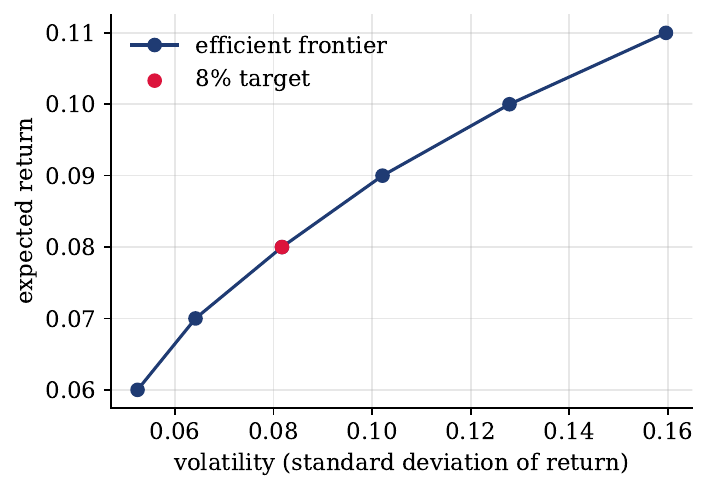}
\caption{Efficient frontier for the mean-variance portfolio QP: expected return against
volatility, swept by a right-hand-side grid on the return-target constraint. The 8\% target
is marked. Recomputed from \code{notebooks/portfolio\_qp.ipynb}.}
\label{fig:frontier}
\end{figure}

The variance rises monotonically as the required return increases, and plotting return against volatility gives the familiar frontier curve.

\textbf{Risk scenarios.} Two risk scenarios exercise the quadratic-scaling experiments and draw a
contrast that the formulation makes precise:

\begin{lstlisting}[style=py]
scenario_analyzer.run_scenarios({
    "base": ScenarioCase(),
    "market_risk_x1.5": ScenarioCase((ScaleQuadraticObjective(1.5),)),
    "variance_surge_x1.5": ScenarioCase((ScaleQuadraticDiagonal(1.5),)),
})
\end{lstlisting}

\begin{center}
\small
\begin{tabular}{lcccccc}
\toprule
scenario & variance & tech & energy & bonds & real\_estate & gold \\
\midrule
base                & 0.0067 & 0.2397 & 0.1063 & 0.1959 & 0.2642 & 0.1938 \\
market\_risk\_x1.5    & 0.0100 & 0.2397 & 0.1063 & 0.1959 & 0.2642 & 0.1938 \\
variance\_surge\_x1.5 & 0.0089 & 0.2141 & 0.1439 & 0.2034 & 0.2732 & 0.1654 \\
\bottomrule
\end{tabular}
\end{center}

A market-wide rise in risk scales the whole covariance and leaves the minimum-variance
portfolio unchanged, rescaling only its variance, because every asset's risk moves by the
same factor. A surge in individual variances, by contrast, scales only the diagonal of the
covariance and rebalances the portfolio toward the steadier assets, because it changes the
assets' relative risk. The two scalings are the \code{ScaleQuadraticObjective} and
\code{ScaleQuadraticDiagonal} experiments of Section~\ref{sec:experiments}, and the contrast between an invariant
portfolio and a rebalanced one is exactly what distinguishes them.

\textbf{The semidefiniteness guard.} The diagonal scaling also shows the guard at work. Scaling
the diagonal upward is always a valid covariance, but scaling it down far enough destroys
positive semidefiniteness and no longer describes a real covariance:

\begin{lstlisting}[style=py]
analyzer.representation.psd_safe_diagonal_scale_lower_bound()   # -> 0.52
analyzer.representation.with_scaled_quadratic_diagonal(0.7)     # accepted
analyzer.representation.with_scaled_quadratic_diagonal(0.5)     # refused
\end{lstlisting}

\pkg{} reports the exact lowest diagonal-scale factor that preserves positive
semidefiniteness (0.52 here) and enforces the exact semidefiniteness check. A scaling of 0.7,
above the factor, is accepted, while a scaling of 0.5, below it, is refused with a message
that cites the factor.

\subsection{One model, five modeling languages}
\label{sec:demo-five}

The demonstrations above build the model in matrix form. Yet, a point of \pkg{}
is that the same analysis surface is recovered no matter how the model was written. This
final demonstration solves the product-mix linear program of Section~\ref{sec:demo-lp} through each
supported modeling language and reads the same answer back.

\textbf{The same answer through five languages.} We express the product-mix model natively in
each of Pyomo, Gurobi, CPLEX, OR-Tools, and cvxpy, wrap it in the matching handle, and
inspect it through \code{linear\_representation()}. While the modeling code changes, everything after the handle is the same. 

In \textbf{Pyomo}, the model is an algebraic
\code{ConcreteModel} and the labor capacity is held as a mutable \code{Param}:

\begin{lstlisting}[style=py]
import pyomo.environ as pyo
from pyoptexplain import Analyzer, PyomoProblemHandle

model = pyo.ConcreteModel()
model.chairs = pyo.Var(domain=pyo.NonNegativeReals)
model.tables = pyo.Var(domain=pyo.NonNegativeReals)
model.labor_cap = pyo.Param(initialize=100.0, mutable=True)
model.labor = pyo.Constraint(expr=2 * model.chairs + model.tables <= model.labor_cap)
model.material = pyo.Constraint(expr=model.chairs + 2 * model.tables <= 80)
model.objective = pyo.Objective(expr=40 * model.chairs + 30 * model.tables, sense=pyo.maximize)

handle = PyomoProblemHandle(model, solver="highs")
\end{lstlisting}

In \textbf{Gurobi}, we build a \code{gp.Model} and tell the handle which variables and constraints map
to the user's named blocks:

\begin{lstlisting}[style=py]
import gurobipy as gp
from pyoptexplain import Analyzer, GurobiProblemHandle

model = gp.Model("production")
model.Params.OutputFlag = 0
chairs = model.addVar(lb=0.0, name="chairs")
tables = model.addVar(lb=0.0, name="tables")
labor = model.addConstr(2 * chairs + tables <= 100, name="labor")
material = model.addConstr(chairs + 2 * tables <= 80, name="material")
model.setObjective(40 * chairs + 30 * tables, gp.GRB.MAXIMIZE)

handle = GurobiProblemHandle(
    model,
    variables={"chairs": chairs, "tables": tables},
    constraints={"labor": labor, "material": material},
)
\end{lstlisting}

In \textbf{CPLEX}, the docplex \code{Model} names its constraints, so the handle reads the mapping back
from the model itself:

\begin{lstlisting}[style=py]
from docplex.mp.model import Model
from pyoptexplain import Analyzer, CplexProblemHandle

model = Model(name="production")
chairs = model.continuous_var(lb=0, name="chairs")
tables = model.continuous_var(lb=0, name="tables")
model.add_constraint(2 * chairs + tables <= 100, ctname="labor")
model.add_constraint(chairs + 2 * tables <= 80, ctname="material")
model.maximize(40 * chairs + 30 * tables)

handle = CplexProblemHandle(model)
\end{lstlisting}

In \textbf{OR-Tools}, the model is a \code{pywraplp} solver built up coefficient by coefficient:

\begin{lstlisting}[style=py]
from ortools.linear_solver import pywraplp
from pyoptexplain import Analyzer, OrToolsProblemHandle

solver = pywraplp.Solver.CreateSolver("GLOP")
chairs = solver.NumVar(0.0, solver.infinity(), "chairs")
tables = solver.NumVar(0.0, solver.infinity(), "tables")

labor = solver.Constraint(-solver.infinity(), 100.0, "labor")
labor.SetCoefficient(chairs, 2.0)
labor.SetCoefficient(tables, 1.0)

material = solver.Constraint(-solver.infinity(), 80.0, "material")
material.SetCoefficient(chairs, 1.0)
material.SetCoefficient(tables, 2.0)

objective = solver.Objective()
objective.SetCoefficient(chairs, 40.0)
objective.SetCoefficient(tables, 30.0)
objective.SetMaximization()

handle = OrToolsProblemHandle(solver)
\end{lstlisting}

In \textbf{cvxpy}, the model is a disciplined-convex \code{Problem}, and the handle is told which
expressions are the user's named variables and constraints:

\begin{lstlisting}[style=py]
import cvxpy as cp
from pyoptexplain import Analyzer, CvxpyProblemHandle

chairs = cp.Variable(nonneg=True, name="chairs")
tables = cp.Variable(nonneg=True, name="tables")
labor = 2 * chairs + tables <= 100
material = chairs + 2 * tables <= 80
problem = cp.Problem(cp.Maximize(40 * chairs + 30 * tables), [labor, material])

handle = CvxpyProblemHandle(
    problem,
    variables={"chairs": chairs, "tables": tables},
    constraints={"labor": labor, "material": material},
)
\end{lstlisting}

From any of the five handles the analysis is identical, and each recovers the optimum 2200
and the same block-facing shadow prices, labor 16.67 and material 6.67:

\begin{lstlisting}[style=py]
analyzer = Analyzer(handle.linear_representation())
analyzer.duals()   # labor 16.6667, material 6.6667 in every case
\end{lstlisting}

What differs is how each scenario study reaches the solver, which is the coverage of Section~\ref{sec:scenarios}.
A native scenario representation mutates the live model in place: Gurobi and CPLEX reach
their own model directly and accept objective-coefficient and variable-bound changes as well
as right-hand-side shifts. Pyomo mutates its algebraic model and additionally supports
reversible in-place constraint ablation. OR-Tools is linear-only through \code{pywraplp}, so its
scenarios route through the structured scenario representation instead, reached with
\code{from\_matrix}. cvxpy registers its scenario variation as \code{cp.Parameter} updates, applied in
place without re-canonicalizing.

\begin{center}
\small
\begin{tabular}{llcl}
\toprule
Language & Handle & Optimum & Scenario path \\
\midrule
Pyomo    & \code{PyomoProblemHandle}    & 2200 & native, structured (matrix) \\
Gurobi   & \code{GurobiProblemHandle}   & 2200 & native, structured (matrix) \\
CPLEX    & \code{CplexProblemHandle}    & 2200 & native, structured (matrix) \\
OR-Tools & \code{OrToolsProblemHandle}  & 2200 & structured (matrix) using \code{from\_matrix} \\
cvxpy    & \code{CvxpyProblemHandle}    & 2200 & native, structured (matrix) \\
\bottomrule
\end{tabular}
\end{center}

For example, the Gurobi native scenario representation applies an objective change, a bound,
and a right-hand-side shift in place:

\begin{lstlisting}[style=py]
Analyzer(handle.scenario_representation()).run_scenarios({
    "base": ScenarioCase(),
    "pricier_chairs": ScenarioCase((SetObjectiveCoefficient("chairs", 55.0),)),
    "cap_chairs": ScenarioCase((SetVariableBounds("chairs", 0.0, 30.0),)),
    "less_material": ScenarioCase((ChangeRHS("material", -20.0),)),
})
\end{lstlisting}

\begin{center}
\small
\begin{tabular}{lcc}
\toprule
scenario & objective\_value & objective\_change \\
\midrule
base           & 2200.0000 & 0.0000    \\
pricier\_chairs & 2800.0000 & 600.0000  \\
cap\_chairs     & 1950.0000 & -250.0000 \\
less\_material  & 2066.6667 & -133.3333 \\
\bottomrule
\end{tabular}
\end{center}

Each scenario is applied to the live Gurobi model and reverted without a rebuild, and the same
study runs unchanged through any of the other four handles along the scenario path its modeling
object supports. The five languages thus reach one common analysis surface, the portability the
architecture of Section~\ref{sec:architecture} is built to provide.

% ===========================================================================
\section{Evaluation}
\label{sec:evaluation}

We ask two questions of \pkg{} in this evaluation:

\begin{itemize}
\item Does it report only what it can justify, and surface backend disagreement instead of
  hiding it (Section~\ref{sec:honesty})?
\item What does the analysis layer cost, on a single analysis (Section~\ref{sec:single-cost}) and at scale
  (Section~\ref{sec:scaling})?
\end{itemize}

Formatting note: throughout, problem sizes are written $m \times n$, with $m$ the number of
constraints and $n$ the number of variables.

The exact code is \pkg{} version 0.1.1, Zenodo DOI: https://doi.org/10.5281/zenodo.21401199.

\subsection{Honesty}
\label{sec:honesty}

Recall the gating principle of Section~\ref{sec:certificates}: \pkg{} reports a quantity only when its
representation certificate justifies it. Two situations
show what this gating buys.

\textbf{Integer duals.} An integer optimum has no dual, but backends differ on what they return
when asked for one. Most guard correctly, by raising or returning \code{None}. Only HiGHS returns
a silent \code{0.0} that a reader could mistake for a shadow price.

\begin{center}
\small
\begin{tabular}{lcccccc}
\toprule
Backend & HiGHS & cvxpy & Pyomo & Gurobi & CPLEX & SCIP \\
\midrule
MILP dual & \code{0.0} & \code{None} & \code{None} & raises & raises & raises \\
\bottomrule
\end{tabular}
\end{center}

\pkg{} gives the same answer on every backend: the integer certificate reports no
dual, and an explicitly derived relaxation supplies an honest valuation on request (on the
generalized-assignment instance, the binding \code{line\_A} capacity is worth $-0.5$ per unit).

\textbf{Degeneracy.} When the optimal basis is not unique, the dual is not unique either:
different solvers settle on different optimal vertices of the same optimal face and so
report different duals for the same problem. This is degeneracy, not solver error. Because
\pkg{} normalizes one representation and hands it to every backend, the disagreement is
easy to read off directly. On the NETLIB~\citep{dongarra1987} instance \code{AFIRO} ($27 \times 32$), the three matrix
backends reach the same optimum ($-464.7531$) but settle on different optimal bases, so their
block duals differ:

\begin{lstlisting}[style=py]
rep = rep_from_mps(netlib_mps("afiro"))          # notebook helper: reads an MPS file
                                                 # into one normalized representation
for backend in (HiGHSBackend, GurobiBackend, CPLEXBackend):
    an = Analyzer(rep, backend=backend())
    an.solve()
    print(backend.__name__, an.duals())           # each backend's own basis-dependent duals
\end{lstlisting}

A representative cluster shows the mechanism: one dual weight is attached to different blocks
by different solvers.

\begin{center}
\small
\begin{tabular}{lccc}
\toprule
Block & HiGHS & Gurobi & CPLEX \\
\midrule
\code{ineq4}           & -2.2704 & -2.2704 & 0 \\
\code{x6\_lower\_bound}  & 2.2704  & 2.2704  & 0 \\
\code{x10\_lower\_bound} & 0       & 0       & 2.2704 \\
\code{ineq9}           & 0       & -2.0922 & 0 \\
\code{x21\_lower\_bound} & 0       & 2.0922  & 0 \\
\code{x25\_lower\_bound} & 2.0922  & 0       & 2.0922 \\
\bottomrule
\end{tabular}
\end{center}

\subsection{Cost of a single analysis}
\label{sec:single-cost}

\textbf{Setup.} The study runs on a single, deliberately modest machine: an Apple M1 with 8
cores and 8\,GB of RAM, under macOS (Darwin 21.6). The interpreter is CPython 3.9.13, which we
report by intent: 3.9 is \pkg{}'s minimum supported Python version, fixed by the oldest
interpreter the optional backends jointly support in this environment, and the study was run on
it. The backends are HiGHS through \code{highspy} and through OR-Tools' \code{pywraplp} (which
bundles its own HiGHS build), Gurobi, CPLEX through \code{docplex}, SCIP
through \code{PySCIPOpt}, OSQP, Clarabel and SCS through cvxpy, and ipopt through Pyomo;
Table~\ref{tab:environment} pins every version. Every timing is wall-clock, in
milliseconds, and is the best of several repetitions to suppress scheduling noise: five
repetitions for the single-analysis costs of this section (ten for the per-front build costs
below), and three for the scenario-scaling tables of Section~\ref{sec:scaling}. Problem data is
generated from fixed seeds, so the instances are deterministic and the structural results
reproduce exactly. Absolute times vary between runs by 10 to 20 percent, so the orderings and
ratios are the result, not the individual milliseconds.

\begin{table}[htbp]
\centering
\caption{Software environment for the computational study. All measurements in
Sections~\ref{sec:single-cost} and~\ref{sec:scaling} use these versions on the machine described
above.}
\label{tab:environment}
\small
\begin{tabular}{llll}
\toprule
Component & Version & Component & Version \\
\midrule
CPython              & 3.9.13     & NumPy    & 1.26.4 \\
\pkg{}               & 0.1.0      & SciPy    & 1.13.1 \\
highspy (HiGHS)      & 1.14.0     & pandas   & 2.3.3 \\
gurobipy (Gurobi)    & 12.0.3     & OSQP     & 1.1.1 \\
docplex (CPLEX 22.1.2.1) & 2.32.264 & cvxpy  & 1.7.5 \\
PySCIPOpt (SCIP)     & 4.2.0      & Clarabel & 0.11.1 \\
OR-Tools (HiGHS 1.7.2) & 9.11.4210  & SCS      & 3.2.11 \\
Pyomo                & 6.9.5      & ipopt    & 3.14.8 \\
\bottomrule
\end{tabular}
\end{table}

This section uses random linear programs $\max\, c^\top x$ s.t.\ $Ax \le b$, $0 \le x \le u$,
with $A$ sparse at density $0.15$, nonzeros $\sim U(0.5, 3)$, $c \sim U(1, 10)$,
$u \sim U(1, 5)$, and each $b_i$ a $U(0.5, 2)$ multiple of row $i$'s coefficient sum plus one.
A solve report is the analysis the user asked for, so the question is not its absolute time
but how it divides among build, solve, and report, and how much the layer adds over the bare
solver.

\textbf{Build, solve, and report.} On the matrix path the representation is the supplied arrays, so
its construction is free and the cost divides between the solve and the report (variables,
block duals, and RHS and objective sensitivity tables).

\begin{center}
\small
\begin{tabular}{cccc}
\toprule
Size ($m \times n$) & Build (ms) & Solve (ms) & Report (ms) \\
\midrule
80 $\times$ 100    & $< 0.01$ & 2.2   & 7.2   \\
400 $\times$ 500   & $< 0.01$ & 22.7  & 35.5  \\
1200 $\times$ 1500 & $< 0.01$ & 218.1 & 171.2 \\
\bottomrule
\end{tabular}
\end{center}

The report falls from about three times the solve at the smallest size to below the solve at
the largest, 3.3x, 1.6x, and 0.78x. Its cost splits across 4
components: primal recovery, duals, and right-hand-side and objective-coefficient ranging. At
the smallest size duals dominate (3.7 of 7.2 ms), while at the largest the three non-primal
components are comparable (46, 60, and 57 ms) with no single component dominant, and primal
recovery is the smallest share throughout (8.4 of 171.2 ms at the largest size).

\textbf{Overhead over the bare solver.} Against a direct \code{highspy} call on the same instances, the
\pkg{} solve path adds a near-fixed cost, so its ratio to the solve falls with size,
from about 4.0x where the solve is sub-millisecond toward one where the solve dominates.

\begin{center}
\small
\begin{tabular}{cccc}
\toprule
Size ($m \times n$) & Bare HiGHS (ms) & \pkg{} (ms) & Ratio \\
\midrule
40 $\times$ 50     & 0.8   & 3.3   & 4.0 \\
150 $\times$ 200   & 3.8   & 9.4   & 2.5 \\
400 $\times$ 500   & 22.4  & 35.7  & 1.6 \\
800 $\times$ 1000  & 105.8 & 128.8 & 1.2 \\
1500 $\times$ 2000 & 370.5 & 386.2 & 1.0 \\
\bottomrule
\end{tabular}
\\[3pt]
{\footnotesize Ratios are computed from the unrounded timings, so they need not equal the
quotient of the two rounded columns.}
\end{center}

\textbf{Build cost by front.} Building a representation from a native modeling object carries that
object's inspection or canonicalization cost. The same two-variable product-mix program built
through each front:

\begin{center}
\small
\begin{tabular}{lcccccc}
\toprule
Front & direct matrix & OR-Tools & CPLEX & Pyomo & Gurobi & cvxpy \\
\midrule
Build (ms) & $< 0.01$ & 0.18 & 0.20 & 0.38 & 0.49 & 1.28 \\
\bottomrule
\end{tabular}
\end{center}

The direct matrix path is free because the arrays are already the representation. cvxpy is the
most expensive because it canonicalizes the model into conic form, the same work that recovers
the block-to-component mapping of Section~\ref{sec:certificates}.

\textbf{Cost by experiment type.} Analyses differ in how many solves they drive. On one 120 $\times$ 150
linear program, with an 8-constraint, 20-variable binary program for the relaxation:

\begin{center}
\small
\begin{tabular}{lcccccc}
\toprule
Experiment & single solve & ablate constr. & ten scenarios & relaxation & perturbation & grid (5$\times$5) \\
\midrule
Time (ms) & 4.0 & 8.5 & 18.2 & 60.1 & 117.8 & 244.4 \\
\bottomrule
\end{tabular}
\end{center}

The cost tracks the number of re-solves. The most expensive analyses re-solve a perturbed
model many times, which raises the scaling question of Section~\ref{sec:scaling}.

\subsection{Scaling repeated scenario analyses}
\label{sec:scaling}

The expensive analyses above re-solve one model under many perturbations. We study the common
case, \textbf{right-hand-side scenarios}: each scenario shifts the RHS of one constraint
by $\delta \sim U(-0.3, 0.3)$, and a batch is $N$ such scenarios on one base program, every one
returning a full solve report.

A scenario representation amortizes work across the batch. We measure three tiers on the same
solver, each returning an identical report, to isolate what the amortization buys:

\begin{itemize}
\item \textbf{naive:} rebuild the solver model and the representation from scratch for every scenario
  and solve cold. This is the user who never adopted a scenario representation.
\item \textbf{re-extract:} build the model once, but rebuild the representation per scenario and solve
  cold with a fresh session.
\item \textbf{reuse:} the scenario representation in full, with one certificate built at extraction and
  a warm solver session reused across cases.
\end{itemize}

Each tier decomposes into three measured phases, \textbf{model} (building the bare solver model or
the matrix arrays), \textbf{cert} (building the representation and its certificate), and \textbf{solve}
(one cold solve including session setup), with $\text{naive} = \text{model} + \text{cert} +
\text{solve}$ and $\text{re-extract} = \text{cert} + \text{solve}$. The reuse tier is the
scenario representation in one of its two forms, introduced in Section~\ref{sec:representations}:

\begin{itemize}
\item the \textbf{structured scenario representation}, \pkg{}'s general path and the
  contribution this study showcases. As explained in Section~\ref{sec:representations}, its lean path edits only the
  changed entries of $b$ on a warm session and reuses the certificate built once at extraction.
  Because the extracted arrays are themselves the model, there is no separate model phase, so
  naive and re-extract coincide. It is available whenever the problem is extractable (linear or
  quadratic objective, linear constraints). Labelled \code{structured} in the tables.
\item the \textbf{native scenario representation}, which mutates the original modeling object in place per
  scenario and restores it. For a solver without its own modeling object it routes through a
  modeling language (cvxpy or Pyomo). Labelled \code{native}.
\end{itemize}

Base programs are random LPs, QPs, MILPs, and MIQPs ($A$ sparse at density $0.25$, nonzeros
$\sim U(0.5, 2)$, one guaranteed nonzero per row, $Q$ diagonal positive definite); the
non-extractable study later in this section adds second-order-cone, exponential-cone, and
general nonlinear programs that no matrix form admits.
Sub-millisecond tiny instances are omitted because their timings are dominated by noise. All figures are milliseconds per scenario.

\subsubsection*{Does reuse pay? Naive versus re-extract versus the scenario representation}

This is the central comparison.

\textbf{Linear programs} ($N = 50$):

\begin{center}
\small
\begin{tabular}{llcrrrrrr}
\toprule
Solver & Path & Size & naive & re-extract & reuse & model & cert & solve \\
\midrule
Gurobi & native     & 200 $\times$ 160 & 38.9  & 32.3  & 3.3   & 6.7  & 22.3  & 9.9   \\
CPLEX  & native     & 200 $\times$ 160 & 83.1  & 48.8  & 4.3   & 34.3 & 6.6   & 42.3  \\
Gurobi & structured & 200 $\times$ 160 & 15.3  & 15.3  & 1.9   & 0.0  & 4.7   & 10.6  \\
CPLEX  & structured & 200 $\times$ 160 & 16.0  & 16.0  & 8.9   & 0.0  & 4.6   & 11.3  \\
HiGHS  & structured & 200 $\times$ 160 & 16.1  & 16.1  & 1.7   & 0.0  & 4.7   & 11.4  \\
Gurobi & native     & 500 $\times$ 400 & 179.5 & 147.2 & 10.1  & 32.3 & 123.2 & 24.0  \\
CPLEX  & native     & 500 $\times$ 400 & 330.5 & 242.2 & 12.9  & 88.3 & 36.1  & 206.1 \\
Gurobi & structured & 500 $\times$ 400 & 58.1  & 58.1  & 5.6   & 0.0  & 12.0  & 46.1  \\
CPLEX  & structured & 500 $\times$ 400 & 56.1  & 56.1  & 40.1  & 0.0  & 11.6  & 44.5  \\
HiGHS  & structured & 500 $\times$ 400 & 63.9  & 63.9  & 4.7   & 0.0  & 11.3  & 52.6  \\
\bottomrule
\end{tabular}
\end{center}

\begin{itemize}
\item Where the solve does not dominate, reuse is far below naive: at 200 $\times$ 160 it cuts the
  per-scenario cost by 12x (Gurobi), 19x (CPLEX), and 9x (HiGHS), and by more at 500 $\times$ 400.
\item The three tiers localize the saving. Re-extract removes the model rebuild (the \code{model}
  column), reuse additionally removes the per-scenario certificate build and the cold-solve
  setup. For native Gurobi the certificate build is the dominant naive cost at scale (123 of
  179 ms at 500 $\times$ 400), and amortizing it once is most of the gain, and for CPLEX the docplex model
  build dominates instead (88 of 330 ms), which re-extract already removes.
\item The structured path is also open to the commercial solvers: routing Gurobi or CPLEX through
  the matrix interface drops the modeling-language build entirely (\code{model} column 0), so even
  their naive cost falls below the native front (Gurobi 15.3 vs 38.9 ms, CPLEX 16.0 vs 83.1 ms at
  200 $\times$ 160), and reuse then strips the certificate as on any structured route.
\end{itemize}

\textbf{Quadratic programs} ($N = 50$):

\begin{center}
\small
\begin{tabular}{llcrrrrrr}
\toprule
Solver & Path & Size & naive & re-extract & reuse & model & cert & solve \\
\midrule
Gurobi & native           & 150 $\times$ 120 & 29.4 & 23.9 & 3.6  & 5.5  & 13.8 & 10.1 \\
CPLEX  & native           & 150 $\times$ 120 & 55.1 & 18.9 & 9.5  & 36.2 & 4.0  & 14.9 \\
OSQP   & native (cvxpy)   & 150 $\times$ 120 & 8.1  & 7.4  & 3.3  & 0.7  & 0.0  & 7.4  \\
Gurobi & structured       & 150 $\times$ 120 & 13.9 & 13.9 & 3.6  & 0.0  & 7.2  & 6.6  \\
CPLEX  & structured       & 150 $\times$ 120 & 18.6 & 18.6 & 7.9  & 0.0  & 7.8  & 10.9 \\
OSQP   & structured       & 150 $\times$ 120 & 22.4 & 22.4 & 2.8  & 0.0  & 7.3  & 15.1 \\
Gurobi & native           & 250 $\times$ 200 & 59.7 & 49.4 & 6.0  & 10.3 & 34.3 & 15.1 \\
CPLEX  & native           & 250 $\times$ 200 & 86.5 & 37.7 & 19.4 & 48.8 & 9.9  & 27.8 \\
OSQP   & native (cvxpy)   & 250 $\times$ 200 & 11.6 & 11.0 & 6.1  & 0.7  & 0.0  & 11.0 \\
Gurobi & structured       & 250 $\times$ 200 & 23.1 & 23.1 & 5.6  & 0.0  & 13.6 & 9.5  \\
CPLEX  & structured       & 250 $\times$ 200 & 28.9 & 28.9 & 10.9 & 0.0  & 14.1 & 14.8 \\
OSQP   & structured       & 250 $\times$ 200 & 53.3 & 53.3 & 4.6  & 0.0  & 13.9 & 39.4 \\
\bottomrule
\end{tabular}
\end{center}

The quadratic case repeats the linear one. Because a right-hand-side scenario leaves the
quadratic objective untouched, neither path re-stages $Q$ per solve, so reuse pays back as
strongly as on an LP (Gurobi native 29.4 to 3.6 ms, CPLEX 55.1 to 9.5 ms). Note that OSQP
 is reached two ways here: the structured representation
drives it directly from the matrix with no modeling-language round-trip (reuse 2.8 and 4.6 ms),
and cvxpy supplies the same solver as a native scenario representation, registering the
right-hand side as a DPP parameter updated per scenario. The two OSQP routes are at parity on
reuse, the same solver reached two ways at the same cost (cvxpy 3.3 and 6.1 ms against the
structured 2.8 and 4.6), because the cvxpy-to-OSQP path is itself a warm-update mechanism that
edits the parameter without re-canonicalizing. Their one-time builds differ in which dominates:
cvxpy's canonicalization is cheaper than the matrix certificate build (a 0.7 ms model build against
a 7.2 ms certificate at 150 $\times$ 120), which is bookkeeping, not solving.

\textbf{Mixed-integer programs} ($N = 30$):

\begin{center}
\small
\begin{tabular}{lllcrrrrrr}
\toprule
Class & Solver & Path & Size & naive & re-extr. & reuse & model & cert & solve \\
\midrule
MILP & Gurobi & native     & 16 $\times$ 12 & 7.5  & 7.2  & 5.5  & 0.3  & 0.4 & 6.8  \\
MILP & CPLEX  & native     & 16 $\times$ 12 & 27.8 & 8.0  & 12.5 & 19.8 & 0.3 & 7.7  \\
MILP & Gurobi & structured & 16 $\times$ 12 & 6.4  & 6.4  & 5.5  & 0.0  & 0.5 & 5.9  \\
MILP & CPLEX  & structured & 16 $\times$ 12 & 19.5 & 19.5 & 13.7 & 0.0  & 0.5 & 19.0 \\
MILP & HiGHS  & structured & 16 $\times$ 12 & 86.9 & 86.9 & 86.2 & 0.0  & 0.5 & 86.3 \\
MILP & SCIP   & structured & 16 $\times$ 12 & 26.0 & 26.0 & 25.4 & 0.0  & 0.5 & 25.5 \\
MIQP & Gurobi & native     & 16 $\times$ 12 & 2.3  & 2.0  & 0.6  & 0.3  & 0.4 & 1.6  \\
MIQP & CPLEX  & native     & 16 $\times$ 12 & 29.0 & 6.1  & 5.1  & 22.9 & 0.3 & 5.8  \\
MIQP & Gurobi & structured & 16 $\times$ 12 & 3.5  & 3.5  & 1.1  & 0.0  & 1.7 & 1.8  \\
MIQP & CPLEX  & structured & 16 $\times$ 12 & 9.1  & 9.1  & 8.6  & 0.0  & 1.1 & 7.9  \\
MIQP & SCIP   & structured & 16 $\times$ 12 & 5.2  & 5.2  & 3.6  & 0.0  & 1.1 & 4.1  \\
\bottomrule
\end{tabular}
\end{center}
Integer programs are solve-dominated: where the branch-and-bound solve is large (about 86 ms for
HiGHS, 26 ms for SCIP) it swamps every other phase, so the three tiers agree and the representation
choice does not matter, the solver sets the cost. On Gurobi and CPLEX the solve is only a few
milliseconds, so the fixed build is a visible share and the small-instance tiers separate and are
noisier: CPLEX's naive cost in particular is inflated by the docplex model build, which re-extract
removes. The seed sweep below shows the invariant that holds regardless, the naive and reuse totals
tracking each other at every $N$ with no crossover.

\textbf{The crossover.} Reuse builds a certificate and a session once, so on a single scenario it can
lose. Sweeping $N$ makes the break-even visible. We showcase this for HiGHS, regenerating the
instance under five seeds at each $N$ and reporting the per-scenario median with its interquartile
range. The speedup is the median of the per-instance naive-to-reuse ratios. Per-scenario cost in
milliseconds:

% Seed-swept N-sweep (gap 5). Reproduce with seed_amortization in notebooks/scenario_study.py
% (median + IQR across seeds=(0..4), n_values=(1,5,20,50,100,200)); update BOTH main.tex and
% oms_paper.tex in the same edit:
%   S.seed_amortization(lambda p: S.matrix_route("LP",   "HiGHS", p, S.HiGHSBackend()),
%                       m=200, n=160, n_values=(1,5,20,50,100,200))
%   S.seed_amortization(lambda p: S.matrix_route("MILP", "HiGHS", p, S.HiGHSBackend()),
%                       m=16, n=12, integer=True, n_values=(1,5,20,50,100,200))
\begin{center}
\small
\begin{tabular}{lrrrr}
\toprule
Class & $N$ & naive (ms) & reuse (ms) & speedup [IQR] \\
\midrule
LP   & 1   & 16.8 & 18.2 & 0.9 [0.9-1.0]    \\
LP   & 5   & 16.5 & 4.7  & 3.5 [3.5-3.6]    \\
LP   & 20  & 16.4 & 2.2  & 7.3 [7.3-7.3]    \\
LP   & 50  & 16.3 & 1.7  & 9.3 [9.3-9.3]    \\
LP   & 100 & 16.4 & 1.6  & 10.4 [10.3-10.4] \\
LP   & 200 & 16.4 & 1.5  & 11.0 [10.7-11.0] \\
MILP & 1   & 39.4 & 40.7 & 1.0 [1.0-1.0]    \\
MILP & 5   & 38.7 & 52.0 & 1.1 [1.0-1.1]    \\
MILP & 20  & 40.1 & 43.9 & 1.1 [1.0-1.1]    \\
MILP & 50  & 39.0 & 37.9 & 1.1 [1.0-1.1]    \\
MILP & 100 & 40.1 & 38.9 & 1.1 [1.0-1.1]    \\
MILP & 200 & 38.9 & 38.3 & 1.0 [1.0-1.1]    \\
\bottomrule
\end{tabular}
\end{center}

\begin{itemize}
\item At $N = 1$ reuse loses (its one-time build is not amortized), so a single what-if is best run
  cold.
\item On a cheap-solve LP it breaks even by about $N = 5$ and climbs past a 10x saving by $N = 100$,
  holding near 11x through $N = 200$.
\item On a solve-dominated MILP there is no crossover: the per-scenario naive and reuse costs track
  each other at every $N$ (median speedup near one, its wide interquartile range reflecting
  instance-to-instance solve variance on the small integer program), and re-solving from scratch is
  as good as any reuse.
\end{itemize}

\subsubsection*{The structured scenario representation versus native representations}

The reuse tier above used, per solver, whichever scenario representation suits it. Here we
compare them head to head. The structured representation always edits the arrays on a warm
session while a native representation reaches the solver three ways, and the route decides its cost:

\begin{itemize}
\item \textbf{direct} (Gurobi, CPLEX): the native object is the solver's own model.
\item \textbf{cvxpy}: registers the right-hand side as a DPP parameter and updates it without
  re-canonicalizing.
\item \textbf{Pyomo}: re-exports the model to the solver on every solve.
\end{itemize}

Reuse-tier cost on a 200 $\times$ 160 LP, $N = 50$:

\begin{center}
\small
\begin{tabular}{lcccc}
\toprule
Solver & Structured representation & Native, direct & Native, cvxpy & Native, Pyomo \\
\midrule
Gurobi & 2.0  & 3.0 & n/a  & n/a   \\
CPLEX  & 8.9  & 4.3 & n/a  & n/a   \\
HiGHS  & 1.8  & n/a & 5.4  & 11.3  \\
SCIP   & 42.0 & n/a & 89.6 & 96.8  \\
\bottomrule
\end{tabular}
\end{center}

\begin{itemize}
\item On Gurobi the structured and direct-native routes are within a factor of two (2.0 vs 3.0 ms),
  both editing a live solver model, so the choice is incidental. On CPLEX the direct-native route is
  the faster by about a factor of two (4.3 vs 8.9 ms), because the structured path re-solves through
  docplex's heavier matrix layer while the native path edits the solver model directly.
\item Where the solver is reached through a modeling language (HiGHS, SCIP), the structured
  representation is the cheapest of these internal routes, because it never re-enters the modeling
  layer per solve. Among the native routes cvxpy's DPP parameters cost less than Pyomo's per-solve
  re-export (on HiGHS, 5.4 vs 11.3 ms), and the structured path is lower still (1.8 ms). The solver
  solves the identical model in the same time on all three, so the spread is the per-solve
  translation the structured representation amortizes once.
\item Guidance: use the structured representation whenever the problem is extractable and reached
  through a modeling language, since there it is fastest and pays no modeling-language round-trip. On
  Gurobi the native and structured routes are interchangeable, on CPLEX the native representation is
  the faster, as its structured matrix path carries the docplex constant. If the model already lives
  in cvxpy its parameter path is a reasonable native option, with Pyomo the slowest.
\end{itemize}

\subsubsection*{Non-extractable models: nonlinear and conic}

The native scenario representation's defining advantage is that it remains available even
when no matrix form exists. We exercise this on programs carrying a second-order-cone,
exponential-cone, or general nonlinear constraint, for which \code{linear\_representation()} and
\code{quadratic\_representation()} refuse extraction outright:

\begin{center}
\small
\begin{tabular}{lll}
\toprule
Backend & Constraint & Matrix extraction \\
\midrule
Gurobi & second-order cone & refused (\code{UnsupportedGurobiStructureError}) \\
CPLEX  & second-order cone & refused (\code{UnsupportedCplexStructureError}) \\
cvxpy  & second-order cone & refused (\code{UnsupportedCvxpyStructureError}) \\
cvxpy  & exponential cone  & refused (\code{UnsupportedCvxpyStructureError}) \\
Pyomo  & nonlinear (\code{exp}) & refused (\code{UnsupportedPyomoStructureError}) \\
\bottomrule
\end{tabular}
\end{center}

The refusal stems from the honesty contract of Section~\ref{sec:analyzer}: the handle raises a typed structure
error and does not attempt any approximation. The native scenario representation then runs the same naive / re-extract / reuse sweep as
before, each backend sweeping the quantity natural to it: Gurobi and CPLEX tighten a linear
budget in a second-order-cone epigraph (an in-place edit on the quadratic block itself is
rejected, so the budget rides a linear row), cvxpy assigns the cone-budget parameter, and Pyomo
with ipopt shifts the nonlinear constraint's right-hand side. 

On a 40-variable instance, $N = 20$, every tier native:

\begin{center}
\small
\begin{tabular}{llrrrrrr}
\toprule
Class & Backend & naive & re-extract & reuse & model & cert & solve \\
\midrule
SOCP     & Gurobi          & 12.2  & 11.9  & 9.9   & 0.4  & 0.6 & 11.3  \\
SOCP     & CPLEX           & 31.3  & 9.3   & 5.8   & 22.0 & 0.2 & 9.1   \\
SOCP     & cvxpy/Clarabel  & 2.8   & 2.5   & 1.7   & 0.3  & 0.0 & 2.5   \\
SOCP     & cvxpy/SCS       & 2.4   & 2.1   & 1.4   & 0.3  & 0.0 & 2.1   \\
exp-cone & cvxpy/Clarabel  & 3.4   & 3.2   & 2.2   & 0.2  & 0.0 & 3.1   \\
NLP      & Pyomo/ipopt     & 270.4 & 270.2 & 270.4 & 0.3  & 0.3 & 269.9 \\
\bottomrule
\end{tabular}
\end{center}

\begin{itemize}
\item Where the build is a real share of the cost, reuse pays back as on the extractable
  programs: the docplex model build dominates the CPLEX naive cost (22 of 31 ms), which
  re-extract and reuse both strip, and the cvxpy fronts shed their canonicalization on reuse
  (Clarabel 2.8 to 1.7 ms, SCS 2.4 to 1.4 ms).
\item The ipopt NLP is solve-dominated: the nonlinear solve is about 270 ms of every tier, so the
  three agree to within the run-to-run noise and the representation choice is immaterial, similar to what we saw previously with MIPs.
\end{itemize}

Note that here a modeling language gets closer to a backend: cvxpy
canonicalizes the cone and hands it to Clarabel or SCS, solvers \pkg{} reaches only through it.

\subsubsection*{Constraint removal: rebuild vs in-place deactivation}

The previous RHS studies were about perturbing a constraint. We now consider the case of removing one altogether. A removal doesn't need to always rebuild the program without the removed constraint: when the constraint is an inequality, the
structured scenario representation instead frees its rows in place, by setting their right-hand
side to infinity, which deactivates the constraint for the optimum while leaving the constraint
matrix and every dimension intact. The deactivation therefore rides the same lean path as a
RHS shift: the certificate is reused rather than rebuilt, and the warm session stays
compatible. We compare two tiers on the right-hand-side problems above, each scenario removing
one inequality block:

\begin{itemize}
\item \textbf{naive:} rebuild the representation with the row physically removed and solve cold.
\item \textbf{deactivation:} free the row in place on the structured scenario representation, with the
  certificate built once and the session reused.
\end{itemize}

The two tiers return the same optimum on every scenario, which we check before timing. Cost per
scenario, $N = 50$:

\begin{center}
\small
\begin{tabular}{lcrrr}
\toprule
Solver & Size ($m \times n$) & naive (ms) & deactivation (ms) & speedup \\
\midrule
HiGHS  & 200 $\times$ 160 & 14.8 & 2.3  & 6.6  \\
HiGHS  & 500 $\times$ 400 & 63.6 & 6.3  & 10.1 \\
Gurobi & 200 $\times$ 160 & 13.2 & 3.3  & 4.0  \\
Gurobi & 500 $\times$ 400 & 56.9 & 9.7  & 5.8  \\
CPLEX  & 200 $\times$ 160 & 15.3 & 8.4  & 1.8  \\
CPLEX  & 500 $\times$ 400 & 60.5 & 40.7 & 1.5  \\
\bottomrule
\end{tabular}
\end{center}

\begin{itemize}
\item Where a warm session is available (HiGHS, Gurobi) the saving matches the right-hand-side study,
  several-fold (about 4 to 10x here), since deactivation amortizes both the certificate build and
  the cold-solve setup.
\item The CPLEX matrix backend has no reusable session, so deactivation amortizes only the
  per-scenario certificate build. The win is smaller and shrinks as the solve comes to dominate.
\item The deactivation is restricted to inequality blocks. An equality cannot be loosened through its
  right-hand side without changing the program's structure, so removing an equality stays on the
  rebuild path.
\item Reporting differs from the ablation experiment of Section~\ref{sec:single-cost} by design: a deactivated row
  still appears in the solve report as inactive, with a zero dual, rather than being dropped from
  the report entirely.
\item An ablation sweep uses the same lean path as a right-hand-side sweep: the
  structured representation frees the inequality in place and reuses the warm session and the
  certificate, so removing a constraint costs no more than shifting one. Modeling languages differ
  on this operation. Regarding modeling languages: Pyomo deactivates a constraint in place through \code{Constraint.deactivate()},
  though whether the following solve avoids a full re-export then depends on the solver interface;
  cvxpy has no in-place equivalent and rebuilds the problem, re-canonicalizing it.
\end{itemize}

\begin{figure}[htbp]
\centering
\includegraphics[width=0.7\textwidth]{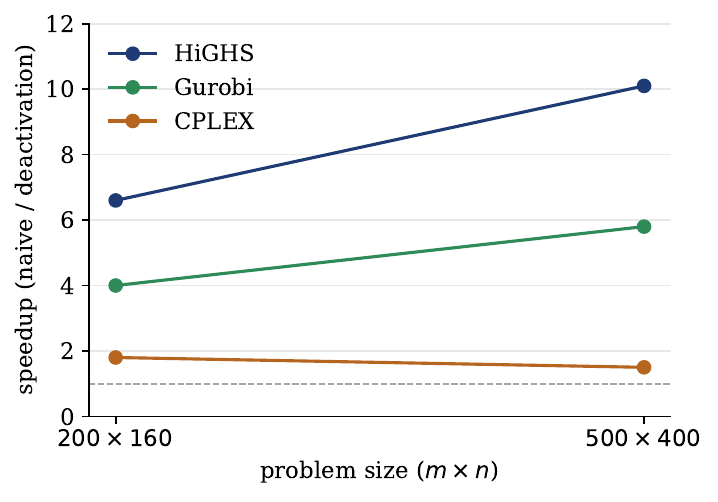}
\caption{Constraint-deactivation speedup against problem size. Freeing an inequality row in
place on the structured scenario representation, against rebuilding the representation without
the row. Where a warm session is available (HiGHS, Gurobi) the saving grows with size to
several-fold; the sessionless CPLEX matrix backend amortizes only the certificate build, and its
win shrinks toward one as the solve comes to dominate. Values are the measured speedups of the
table above ($N = 50$).}
\label{fig:deactivation}
\end{figure}

\subsubsection*{An external baseline: practitioner workflows}

The comparisons above are all internal to \pkg{}. They establish that reuse is efficient, but
not that the integration is worth adopting over how a practitioner gets this information today.
We close the section by placing the reuse path against the routes that do not use \pkg{} at
all, on the same right-hand-side sweep, with every route checked to return the reference
objective on every scenario. The solver is held fixed across each comparison so that the result
is a property of the workflow and not of the solver: HiGHS for the linear sweep and OSQP for the
quadratic one. The routes are:

\begin{itemize}
\item \textbf{direct solver library (the floor)}: a hand-written loop calling the bare solver API
  (\code{highspy} for the LP, \code{osqp} for the QP, with \code{gurobipy} and \code{docplex} for
  context), editing the right-hand side and re-optimizing the warm model in place. It produces a
  bare objective and no diagnostics. Because \pkg{} uses the solver rather than replacing it,
  this is the genuine floor, and the most the reuse path can do is approach it.
\item \textbf{modeling language}: the same model driven repeatedly from a modeling language. A
  Pyomo model re-exported to the solver on every solve, an OR-Tools \code{pywraplp} model whose
  bound is edited in place but which \code{pywraplp} re-exports to the solver on every
  \code{Solve}, and a cvxpy problem driven by a DPP \code{Parameter}, which alone among the three
  updates the right-hand side without re-canonicalizing. This is the repeated-solve workflow a
  practitioner writes when they want more than a single number and stay inside the modeling
  language.
\item \textbf{\pkg{} reuse}: the structured matrix scenario representation, which takes the model
  through the same modeling-language entry point once, extracts the matrix, and then reuses it
  across the sweep on a warm solver session.
\end{itemize}

The design intent of Section~\ref{sec:design} is visible directly in this comparison. The
objective of \pkg{} here is not speed but convenience: one uniform interface over the
modeling-language and solver landscape, on which post-optimality and what-if analyses are easy and
consistent to run. The efficiency result is the supporting one, that adopting the layer costs the
practitioner essentially nothing over the workflow they already use. \pkg{} keeps the modeling
language as the entry point, the part it is good at, namely turning the user's formulation into
solver form, and pays that cost once; it then drives the solver across the sweep on a warm session.
We do not set out to out-solve anyone: the bare solver is the floor, and the modeling-language
routes are the relevant point of comparison. Against them the structured path matches or improves
on the current workflow, faster on linear programs and comparable on quadratic ones, while on every
scenario it returns far more than a re-solve, namely the full certified report and the block
structure recovered at the entry point. Linear programs, per scenario, $N = 50$:

\begin{center}
\small
\begin{tabular}{lllrr}
\toprule
Route & Tool & Solver & 200$\times$160 & 500$\times$400 \\
\midrule
direct solver library (floor) & highspy                   & HiGHS  & 0.3 & 1.6  \\
direct solver library (floor) & gurobipy                  & Gurobi & 0.4 & 2.1  \\
direct solver library (floor) & docplex                   & CPLEX  & 1.3 & 3.4  \\
native multi-scenario         & gurobipy (multi-scenario) & Gurobi & 0.4 & 1.9  \\
modeling language             & cvxpy (Parameter)         & HiGHS  & 4.1 & 14.7 \\
modeling language             & Pyomo (re-export)         & HiGHS  & 4.0 & 13.3 \\
modeling language             & OR-Tools (re-export)      & HiGHS  & 7.2 & 41.2 \\
\pkg{} reuse                  & structured                & HiGHS  & 1.8 & 4.6  \\
\pkg{} reuse                  & structured                & Gurobi & 1.8 & 5.2  \\
\pkg{} reuse                  & structured                & CPLEX  & 8.7 & 37.6 \\
\bottomrule
\end{tabular}
\end{center}

\begin{itemize}
\item On the solvers with a lean update (HiGHS, Gurobi) the reuse path stays within a small
  factor of the bare floor, and the gap narrows with size as the solve amortizes the fixed
  bookkeeping: against \code{highspy} the structured path is $1.6$ to $4.6$ ms at 500 $\times$
  400 ($2.9\times$), against \code{gurobipy} $2.1$ to $5.2$ ms ($2.5\times$), tighter than at
  200 $\times$ 160 where the fixed build is a larger share.
\item Gurobi's native multi-scenario facility (the \code{native multi-scenario} row) sits at the
  floor, $0.4$ and $1.9$ ms, marginally under its own sequential \code{gurobipy} loop ($0.4$ and
  $2.1$ ms), since on a right-hand-side sweep a warm re-solve is already near-minimal and the batch
  saves only per-call overhead. It returns objectives alone; the contrast with the reuse path is
  drawn at the end of this section.
\item On linear programs the structured path is faster than the modeling-language routes that
  re-translate per solve. On HiGHS it costs $4.6$ ms at 500 $\times$ 400 against cvxpy's $14.7$,
  Pyomo's $13.3$, and OR-Tools' $41.2$ (and $1.8$ against $4.1$, $4.0$, and $7.2$ at 200 $\times$
  160), because it never re-enters the modeling layer per solve, the same per-solve translation
  isolated in the matrix-versus-native study above. OR-Tools illustrates that per-solve cost: it
  carries no native scenario representation in \pkg{} and is matrix-only, so a practitioner staying
  inside it edits the bound in place yet \code{pywraplp} still re-exports the whole model to HiGHS
  on every \code{Solve}, exactly the round-trip the matrix path extracts once and then discards.
  (OR-Tools drives its bundled HiGHS, the same solver at a different patch level than the
  \code{highspy} floor.)
\item CPLEX carries a larger constant on both the floor and the reuse path: its \code{docplex}
  floor is the slowest of the three ($3.4$ ms at 500 $\times$ 400), and its matrix scenario path
  re-solves through docplex's heavier Python layer, so the structured reuse sits at $37.6$ ms. The
  pattern, not the constant, is the result.
\end{itemize}

The quadratic sweep repeats the comparison on OSQP, the open-source QP solver, with the floor the
bare \code{osqp} library. Two of the modeling-language routes cannot reach OSQP, and we exclude
them rather than swap the solver: Pyomo exposes no OSQP interface in this environment, and
OR-Tools' \code{pywraplp} is linear-only. The cvxpy DPP-parameter route does reach OSQP and is the
one modeling-language baseline that survives. Per scenario, $N = 50$:

\begin{center}
\small
\begin{tabular}{lllrr}
\toprule
Route & Tool & Solver & 200$\times$160 & 350$\times$200 \\
\midrule
direct solver library (floor) & osqp              & OSQP   & 1.1  & 2.1  \\
modeling language             & cvxpy (Parameter) & OSQP   & 3.0  & 4.6  \\
\pkg{} reuse                  & structured        & OSQP   & 3.5  & 5.1  \\
\pkg{} reuse                  & structured        & Gurobi & 4.9  & 6.3  \\
\pkg{} reuse                  & structured        & CPLEX  & 9.8  & 12.3 \\
\bottomrule
\end{tabular}
\end{center}

\begin{itemize}
\item Against the \code{osqp} floor the reuse path behaves as on the linear sweep, within a
  factor and closing with size: $1.1$ to $3.5$ ms at 200 $\times$ 160 and $2.1$ to $5.1$ ms at
  350 $\times$ 200.
\item Against the cvxpy parameter route the two are very close: cvxpy at $3.0$ and $4.6$ ms, the
  structured path at $3.5$ and $5.1$ ms. This is expected, because cvxpy's DPP parameters are
  themselves a reuse mechanism, updating the right-hand side without re-canonicalizing, so neither
  route re-translates per solve and the timings track each other. The same parity appears whether cvxpy is
  used outside \pkg{} as this baseline or adopted inside it: the QP tier table above also exposes
  cvxpy on OSQP as a native scenario representation, and its reuse cost tracks the structured OSQP
  path for the same reason.
\item The commercial QP backends are reported for context: Gurobi tracks OSQP closely ($4.9$ and
  $6.3$ ms), while CPLEX again carries its heavier Python constant ($9.8$ and $12.3$ ms).
\end{itemize}

On both sweeps the reuse path approaches the bare floor from above and never undercuts it,
because \pkg{} is a layer above the solver and not a solver. The gain here is in what each scenario returns: the floor
gives an objective and nothing else, whereas \pkg{} returns the full certified solve report, the
duals, the sensitivity tables, and the block-to-component mapping, on every scenario, for a small
constant over the floor and comparable to or below the modeling-language routes. The same layer folds structured operations into
that amortized path, namely in-place constraint deactivation: a cvxpy workflow rebuilds and re-canonicalizes to drop a constraint, and while Pyomo can
deactivate one in place its solve may still re-export, whereas the structured scenario representations free
the inequality in place and reuses the warm session and certificate uniformly, so an ablation sweep
runs at the same lean cost as a right-hand-side sweep.

Note: Gurobi does provide a purpose-built multi-scenario facility (\code{NumScenarios} with per-scenario
\code{ScenN*} attributes), which loads a batch of objective, bound, and right-hand-side changes
and solves them in one pass. The \code{native multi-scenario} row measures it at $0.4$ and $1.9$
ms per scenario, at the bare-solver floor. The \pkg{} reuse path is about three to five times
slower ($1.8$ and $5.2$ ms on Gurobi), the price of a certified per-scenario report rather than
objectives alone, and of the same interface across the other four modeling languages and backends.
This native functionality is the better choice for a batch of objectives on Gurobi, while \pkg{} is the better choice
when the analysis and its certificate must hold regardless of the modeling language or backend.

\subsubsection*{Summary}

\begin{center}
\small
\renewcommand{\arraystretch}{1.2}
\begin{tabularx}{\textwidth}{lX}
\toprule
Question & Answer \\
\midrule
Is a scenario representation worth adopting? & Yes once $N$ exceeds a handful, except where the solve dominates \\
What does the reuse amortize? & The per-scenario model build, certificate build, and cold-solve setup. For native solvers the certificate build is the largest of these at scale \\
When does it not help? & Single what-ifs ($N = 1$) and solve-dominated problems (hard integer, large SCIP), where the solver sets the cost \\
Structured or native, commercial solver? & Gurobi: interchangeable, within a factor of two. CPLEX: the native route is faster, its structured matrix path carrying the docplex constant \\
Structured or native, open-source solver? & Structured, to avoid the modeling-language round-trip and amortize the bookkeeping operations\\
Nonlinear or conic (non-extractable)? & Native scenario representation only: the matrix path is refused, and reuse still amortizes the build where the solve does not dominate \\
Worth it over a manual workflow? & Adopting it costs essentially no efficiency: within a constant factor of the bare-solver floor and matching or improving on the modeling-language routes. The gain is the uniform interface, the full certified report on every scenario, and structured operations such as in-place ablation folded into the same amortized path across every backend \\
\bottomrule
\end{tabularx}
\end{center}

% ===========================================================================
\section{Conclusion and future work}
\label{sec:conclusion}

\pkg{} is a post-optimality analysis layer for optimization models built in Python. It
adapts a model authored in any of several modeling languages into a normalized internal
representation, solves it through a choice of backends, and reports the post-optimality
quantities that the representation and the backend can justify, declining the rest rather than
approximating them. It also answers the what-if questions practitioners raise through scenario analysis, reusing a warm analysis surface so repeated queries stay close to the bare-solver cost. We presented the design behind it, the
representation certificate and capability gating that enforce it, the analyses it supports from
the solve report through perturbation and scenario analysis, and a computational study of its
honesty and of what the analysis layer costs at scale. Our aim throughout has been a
practical one: to make the why and what-if questions of an optimization decision easy to ask
and to answer, uniformly, across the tools practitioners already use. The computational study
makes the position concrete. Because \pkg{} uses the solvers rather than replacing them, the
bare solver is the floor it works against, not a target to beat. Against a repeated-solve
modeling-language workflow it is faster on linear programs and comparable on quadratic ones, since
it pays the canonicalization once at the entry point rather than on every solve; the point of the
layer, though, is less the timing than what each scenario returns, namely the full certified
post-optimality report and the recovered structure rather than a bare re-solve.

These design choices are also a foundation for further work, and
we see two natural directions:

\begin{itemize}
\item \textbf{Broader coverage.} The library can grow along its existing axes: further modeling languages
  as front ends, further solver backends, additional capabilities on the backends already
  integrated, and structured representations for model classes beyond the linear and quadratic
  forms supported today.
\item \textbf{Explanation methods.} A certified post-optimality layer is a natural base for explanations
  that go beyond duals and ranges, in particular Shapley-value attribution of an objective or a
  decision to the constraints and parameters that drive it, with the efficient computation that
  realistic models demand \citep{shapley1953, le2020, tsuchiya2023}, and counterfactual explanation, which answers a what-if question constructively by
  finding the smallest change to the data that produces a desired solution \citep{kurtz2026}.
\end{itemize}

Each of these fits the same principle that governs \pkg{}, namely that an analysis is
offered only when it can be justified. We designed the normalized representation to make such
additions incremental rather than disruptive.

% ===========================================================================
\bibliography{references}

\end{document}